\documentclass[fleqn,usenatbib]{mnras}

\usepackage{caption}
\usepackage{subcaption}
\usepackage{color,soul}
\usepackage{hyperref}
\usepackage{comment}

\usepackage{newtxtext,newtxmath}
\usepackage[T1]{fontenc}
\usepackage[fleqn]{nccmath}

\DeclareRobustCommand{\VAN}[3]{#2}
\let\VANthebibliography\thebibliography
\def\thebibliography{\DeclareRobustCommand{\VAN}[3]{##3}\VANthebibliography}

\usepackage{graphicx}	% Including figure files
\usepackage{amsmath}	% Advanced maths commands
\usepackage[flushleft]{threeparttable}
\title[Strong lensing constraints on ULDM]{Quantum fluctuations masquerade as halos: Bounds on ultra-light dark matter from quadruply-imaged quasars}

\author[A. Laroche et al.]{
Alexander Laroche,$^{1,2}$\thanks{E-mail: alexander.laroche@mail.mcgill.ca}
Daniel Gilman,$^{1}$\thanks{E-mail: gilman@astro.utoronto.ca}
Xinyu Li,$^{3, 4}$
Jo Bovy,$^{1}$
Xiaolong Du$^{5}$
\\
$^{1}$Department of Astronomy and Astrophysics, University of Toronto, Toronto, ON, M5S 3H4, Canada\\
$^{2}$Department of Physics, McGill University, Montreal, QC, H3A 2T8, Canada\\
$^{3}$ Canadian Institute for Theoretical Astrophysics, 60 St George St, Toronto, ON M5R 2M8\\
$^{4}$Perimeter Institute for Theoretical Physics, 31 Caroline Street North, Waterloo, Ontario, Canada, N2L 2Y5\\
$^{5}$Carnegie Institution for Science, 813 Santa Barbara Street, Pasadena, CA 91101, USA
}

\date{Accepted . Received}

\pubyear{2022}

\begin{document}
\label{firstpage}
\pagerange{\pageref{firstpage}--\pageref{lastpage}}
\maketitle

\begin{abstract}
Ultra-light dark matter (ULDM) refers to a class of theories, including ultra-light axions, in which particles with mass $m_{\psi} < 10^{-20}\, \rm{eV}$ comprise a significant fraction of the dark matter. A galactic scale de Broglie wavelength distinguishes these theories from cold dark matter (CDM), suppressing the overall abundance of structure on sub-galactic scales, and producing wave-like interference phenomena in the density profiles of halos. With the aim of constraining the particle mass, we analyze the flux ratios in a sample of eleven quadruple-image strong gravitational lenses. We account for the suppression of the halo mass function and concentration-mass relation predicted by ULDM theories, and the wave-like fluctuations in the host halo density profile, calibrating the model for the wave interference against numerical simulations of galactic-scale halos. We show that the granular structure of halo density profiles, in particular, the amplitude of the fluctuations, significantly impacts image flux ratios, and therefore inferences on the particle mass derived from these data. We infer relative likelihoods of CDM to ULDM of 8:1, 7:1, 6:1, and 4:1 for particle masses $\log_{10}(m_\psi/\rm{eV})\in[-22.5,-22.25], [-22.25,-22.0],[-22.0,-21.75], [-21.75,-21.5]$, respectively. Repeating the analysis and omitting fluctuations associated with the wave interference effects, we obtain relative likelihoods of CDM to ULDM with a particle mass in the same ranges of 98:1, 48:1, 26:1 and 18:1, highlighting the significant perturbation to image flux ratios associated with the fluctuations. Nevertheless, our results disfavor the lightest particle masses with $m_{\psi} < 10^{-21.5}\,\rm{eV}$, adding to mounting pressure on ultra-light axions as a viable dark matter candidate. 
\end{abstract}


%%%%%%%%%%%%%%%%%%%%%%%%%%%%%%%%%%%%%%%%%%%%%%%%%%

%%%%%%%%%%%%%%%%% BODY OF PAPER %%%%%%%%%%%%%%%%%%

\section{Introduction}

In recent decades, numerous alternative theories have emerged that challenge the reigning cosmological model of cold dark matter (CDM). One such theory, referred to as ultra-light dark matter (ULDM), predicts that dark matter is composed of light bosons with particle masses $m_{\psi} < 10^{-20}\,\rm{eV}$. Light axions are a particularly well-motivated ULDM particle candidate that can arise from string theories, with a canonical mass $m_\psi\sim10^{-22}$ eV \citep{Witten1984,Svrcek2006,Amendola06,arvanitaki2010,acharya2010,cicoli2012,Cicoli++21}. Early motivation for ultra-light axions came after various authors pointed out that axion-like particles could potentially resolve the strong CP problem in quantum chromodynamics \citep{peccei1977,weinberg1978,wilczek1978,Preskill1982,Abbott1982,Dine1982}.

Encouraged by the utility of ULDM theories as elegant resolutions to longstanding challenges in particle physics, attention turned to astrophysical implications. The possibility that ULDM can resolve small-scale challenges to $\Lambda$CDM (for a review, see \citet{Bullock2017}) given $m_{\psi} \sim 10^{-22}\,\rm{eV}$ \citep{Hu2000} provides impetus to investigate the consequences of ULDM theories on structure formation on sub-galactic scales. The de Broglie wavelength associated with the particle mass sets the physical scale relevant for cosmic structure formation. For a particle mass $m_{\psi} = 10^{-22}\,\rm{eV}$ and a typical virial velocity for a galaxy of $\sim 100\,\rm{km} \ \rm{s^{-1}}$, the de Broglie wavelength is on the order of one kilo-parsec (kpc). Hence, quantum mechanical phenomena manifest on length scales comparable to the size of a galaxy. 

The kpc-scale de Broglie wavelength predicted by ULDM theories has three main consequences for structure formation in the Universe. First, inside individual halos, quantum pressure between the particles, sometimes phrased in terms of the uncertainty principle, leads to the formation of a flat central region referred to as a soliton core \citep{Schive2014a}. For this reason, ULDM has been proposed as a potential solution to the so-called core-cusp problem \citep[e.g.][]{Kendall++20}. Second, quantum pressure precludes the collapse of small-scale density fluctuations in the early Universe, leading to a cutoff in the matter power spectrum below a characteristic scale related to the particle mass. This cutoff suppresses both the abundance and concentration of dark matter halos below a certain halo mass scale, relative to CDM \citep{Schive2014a,Du2016,Schive2016,Kulkarni_2021}. Third, quantum mechanical effects give rise to wave-like interference patterns in halo density profiles, causing fluctuations in the mass density with a typical size comparable to the de Broglie wavelength and an amplitude proportional to the average local density
\citep[e.g.][]{Magana2012,Suarez2013,Hui2017,Li++2019,ferreira2021ultralight,Li++2021,Yavetz++22}. If present, these fluctuations could add an additional source of small-scale density fluctuation, in addition to subhalos, to the mass profile of galactic halos.  

Existing constraints on ULDM come from a variety of cosmological probes on vastly different physical scales. Comparisons between structure formation in simulations with ULDM and the Lyman-$\alpha$ forest yield a 95\% lower limit $m_\psi>2\times10^{-20}$ eV \citep{Keir2021}. \citet{Dalal++22} claimed a constraint $m_{\psi}>3\times10^{-19}\rm{eV}$ at $99\%$ confidence based on stellar dynamics in ultra-faint dwarf galaxies. \citet{Davoudiasl_Denton19} have disfavored $m_\psi\sim10^{-21}$ eV through observations of the Messier 87 super-massive black hole, via superradiance. Motivated by the core-cusp problem, using Jeans analysis \citet{Chen++2017} were able to explain the flat central density profiles in dwarf spheroidal (dSph) galaxies with ULDM, provided that $m_\psi\sim10^{-22}$ eV. dSphs have also been used by \citet{MarshPop15} and \citet{Gonzalez-Morales++17} to obtain upper bounds $m_\psi<1.1(0.4)\times10^{-22}$ at $95(97.5)\%$ confidence, respectively. When taking all of these constraints into account, there is an apparent tension between boson mass lower bounds from, for instance, the Lyman-$\alpha$ forest and stellar dynamics in UFDs and upper bounds from dSphs. The constraints on the particle mass from the Lyman-$\alpha$ forest, stellar dynamics in ultra-faint dwarfs, black hole superradiance and dSphs depend on indirect probes of dark matter structure, in the sense that the observable used to constrain the particle mass is associated with baryonic physics. \citet{Banik++21} directly constrained the ULDM subhalo mass function with measurements of the Milky Way's subhalo mass function using stellar streams from \citet{Banik++21a}, finding $m_\psi > 2.2 \times 10^{-21}$ eV at 95$\%$ confidence. This confirmed prior work by \citet{Schutz2020}, who translated inferences on the free-streaming length of warm dark matter (WDM) presented by \citet{Gilman2019a} and \citet{Banik++21} into constraints on the ULDM particle mass, concluding that $m_\psi>2.1\times10^{-21}$ eV at $95 \%$ confidence. The constraints presented by \citet{Schutz2020} assume that the mapping from WDM to ULDM produces weaker constraints on the ULDM particle mass than one would obtain when accounting for all of the relevant physics that distinguishes the two models. While \citet{Schutz2020} partially based their constraints on the results of a strong lensing analysis presented by \citet{Gilman2019a}, the strong lensing analysis examined WDM, not ULDM. Thus, the constraints presented by \citet{Schutz2020} do not account for the full suite of physics that distinguishes WDM from ULDM. The full power of strong lensing as a tool to study ULDM remains unexplored. 

Over the past few decades, strong lensing has matured as an effective method to constrain dark matter models on sub-galactic scales, below $10^{10}\,M_\odot$ \citep{Dalal2002,Vegetti2014,Nierenberg2014,Hezaveh2016,Inoue16,Nierenberg2017,Birrer2017,Ritondale++19,Cyr-Racine++19,Hsueh2019,Gilman2019a,Gilman2019c,Gilman2021,Gilman2022,He++20,Despali++22,Wagner-Carena++22,Dhanasingham++22}. 
As a purely gravitational phenomena, strong lensing can probe populations of dark matter halos without relying on stellar mass or other luminous material as a tracer for dark matter. Lensing provides a completely independent means with which to test the predictions of ULDM from analyses based on stellar streams, and provides a more direct probe of dark matter structure than the Lyman-$\alpha$ forest, and inferences based on dwarf galaxies and stellar dynamics.

The relative brightness (flux ratios) between images in quadruple-image lenses (quads) are extremely sensitive to the presence of low-mass structures, typically associated with low-mass halos, near lensed images. These halos can either exist around the main deflector as subhalos of a more massive host, or along the line of sight between the observer, main deflector, and source. The abundance and density profiles of subhalos and line-of-sight halos impact the flux ratios, and as such, these data can test any dark matter theory that alters the overall abundance or density profile of halos \citep[e.g.][]{Banik++19,Vegetti++18,Hsueh2019,Despali++20,Gilman2019a,Gilman2021,Minor++21,Amorisco++22,Zelko++22}, as well as the primordial matter power spectrum that set the initial conditions for structure formation \citep{Gilman2022}. Existing analyses of quads use these data to detect individual halos \citep{Nierenberg2014,Nierenberg2017}, and to study entire populations of objects \citep[e.g.][]{Hsueh2019,Gilman2019a,Gilman2019b,Gilman2022}. With regard to ULDM, \citet{Chan2020} (see also \citet{kawai2022}) point out that the quantum fluctuations of the host dark matter halo profile can impart measurable perturbation to image flux ratios. This feature sets ULDM apart from other dark matter frameworks, where only halo abundance and density profiles distinguish a theory from CDM.

In this work, we constrain ULDM models with a sample of eleven quadruply-imaged quasars using accurate models for the halo mass function, concentration-mass relation, soliton cores, and the quantum density fluctuations of the host halo, all of which can affect strong lensing flux ratios and therefore the constraints on the ULDM particle mass derived from them. The structure formation model we implement for the halo mass function is calibrated from numerical simulations \citep{Schive2016}. We derive a new concentration-mass relation for ULDM with precise forms of the transfer function that alters the linear matter power spectrum using the semi-analytic model {\tt{galacticus}} \citep{Benson12} and methods discussed by \citet{Schneider++15}. Finally, using numerical methods presented by \citet{Yavetz++22}, we simulate host dark matter halos with virial masses of $\sim 10^{13} M_{\odot}$ to implement a model for the fluctuations of the host dark matter density profile. Combining each of these modeling ingredients, we use the Bayesian inference framework developed and tested by \citet{Gilman2019c,Gilman2019a} to obtain constraints on the mass of the ultra-light dark matter particle, and examine the contribution of each component (fluctuations only, halos only, and the full model including both) to the signal extracted from the data. The data we use to perform this analysis is presented by \citet{chiba2005,Sugai07,Nierenberg2014,Nierenberg2017,Stacey18,Nierenberg2019}, and consists mainly of image fluxes measured from the nuclear narrow-line emission around the background quasar. These data have source sizes compact enough to retain sensitivity to milli-lensing by dark matter halos, but large enough that they are immune to micro-lensing by stars, and variability of the background quasar on scales less than the light crossing time of the emission region. 

This paper is organized as follows. In Section \ref{sec:inferencemethod}, we describe the Bayesian inference methodology. In Section \ref{sec:model}, we discuss the modeling of substructure in ULDM, including halo mass profiles, subhalo and field halo mass functions, the mass-concentration relation, and quantum fluctuations caused by the wave interference effects. In Section \ref{sec:summary_stat}, we examine the effect of ULDM substructure and fluctuations on flux ratios. We present our main results in Section \ref{sec:results}, and give concluding remarks in Section \ref{sec:discussion}.

We assume values for cosmological parameters for flat $\Lambda$CDM measured by Planck \citet{Planck2020}. Lensing computations are performed using the open source gravitational lensing software \texttt{lenstronomy}\footnote{https://github.com/sibirrer/lenstronomy} \citep{Birrer2018,Birrer2021}. Subhalo and line of sight halo populations, as well as the model for the quantum fluctuations in the halo density profile, are implemented in the open source software \texttt{pyHalo}\footnote{https://github.com/dangilman/pyHalo}. Flux ratio forward modeling was performed using \texttt{quadmodel}\footnote{https://github.com/dangilman/quadmodel}. We use \texttt{galacticus}\footnote{https://github.com/galacticusorg/galacticus} \citep{Benson12} to compute the concentration-mass relation for ULDM. We assume a halo mass definition calculated with respect to 200 times the critical density of the Universe at the halo redshift. The mass and density profiles of subhalos are defined with respect to this convention by evaluating the critical density at the time of infall. 

\section{Inference method and dataset}
\label{sec:inferencemethod}

In this section, we review the inference methodology used in this work, and the data we analyze with it. Section \ref{ssec:inferencemethodology} discusses how we forward model the data to evaluate the likelihood function, and Section \ref{ssec:data} discusses the dataset and factors that determine the sample selection. Additional discussion regarding the methodology used in this work are discussed in Section 2 in \citet{Gilman2022}, and Appendix C in \cite{Gilman2019c} gives additional details regarding the ray-tracing algorithms. 

\subsection{Inference method}
\label{ssec:inferencemethodology}
The inference method in this work follows the methodology detailed in \citet{Gilman2019c,Gilman2019a}.
Our goal in this work is to compute the posterior probability distribution of a set of hyper-parameters, ${\bf{q}}_{\rm{sub}}$, that define properties of ultra-light dark matter, given a sample of quadruply-imaged quasars with data ${\bf{D}}$. Labelling the data for the $n$th lens $\bf{d}_n$, as each lens system contributes an independent source of information, we can express the posterior as a product of individual likelihoods
\begin{ceqn}
\begin{equation}
\label{eqn:posterior}
    p\left({\bf{q}}_{\rm{sub}} | {\bf{D}}\right) \propto \pi\left({\bf{q}}_{\rm{sub}}\right) \prod_n \mathcal{L}\left(\bf{d}_n | \bf{q}_{\rm{sub}} \right),
\end{equation}
\end{ceqn}
where $\pi\left(\bf{q}_{\rm{sub}}\right)$ is the prior on the hyper-parameters, and $\mathcal{L}\left({\bf{d}}_n | {\bf{q}}_{\rm{sub}} \right)$ is the likelihood of the $n$th dataset given the parameters.

We compute the individual likelihoods by generating realizations, $\bf{m}_{\rm{sub}}$, of dark matter halos from the model specified by $\bf{q}_{\rm{sub}}$ using $\tt{pyHalo}$. A single realization defines the positions, masses, density profiles, and redshifts of halos between the observer and the source that can impact the data. The likelihood also includes a marginalization over a set of nuisance parameters, $\bf{x}$, which include the size of the lensed background source, and parameters that describe the mass profile of the main deflector
\begin{ceqn}
\begin{equation}
\label{eqn:likelihood}
    \mathcal{L}\left(\bf{d}_n | \bf{q}_{\rm{sub}}\right) = \int p\left(\bf{d}_n | \bf{m}_{\rm{sub}}, \bf{x}\right) p\left(\bf{m}_{\rm{sub}}, \bf{x} | \bf{q}_{\rm{sub}}\right) d \bf{m}_{\rm{sub}} d \bf{x}.
\end{equation}
\end{ceqn}

To evaluate this integral, we use the forward modeling methodology developed and tested by \citet{Gilman2019c,Gilman2019a}. First, using the software package {\tt{pyHalo}}, we generate $\bf{m}_{\rm{sub}}$ from the model specified by $\bf{q}_{\rm{sub}}$. Next, we use the multi-plane lens equation \citep{BlanfordNarayan86} 
\begin{ceqn}
	\begin{equation}
	\label{eqn:raytracing}
	\boldsymbol{\theta_K} = \boldsymbol{\theta} - \frac{1}{D_{\rm{s}}} \sum_{k=1}^{K-1} D_{\rm{ks}}{\boldsymbol{\alpha_{\rm{k}}}} \left(D_{\rm{k}} \boldsymbol{\theta_{\rm{k}}}\right)
	\end{equation}
\end{ceqn} 
to map the four image positions of the lensed quasar to a common source position, with the halos specified by $\bf{m}_{\rm{sub}}$ included in the computation. The notation $D_{ij}$ represents an angular diameter distance between the $i$th and $j$th lens plane, while $D_{\rm{s}}$ represents the angular diameter distance to the source plane. The vector $\boldsymbol{\theta}$ represents an angle on the sky. $\boldsymbol{\alpha_{\rm{k}}}$ and $D_{k}$ are the deflection field and the angular diameter distance at the $k$th lens plane, respectively.

In this step, we also specify an analytic model for the main deflector mass profile, which comprises the smoothly-distributed\footnote{By smoothly-distributed, we mean that the model does not contain large fluctuations in the projected mass on angular scales comparable to the size of a lensed image.} background density of the host dark matter halo, and the luminous galaxy at its center. Main deflector galaxies are typically massive ellipticals embedded in external shear \citep{Gavazzi++07,Auger++10}, so we model the main deflector as a power-law ellipsoid plus external shear. We add radial flexibility to this profile by sampling different logarithmic profile slopes. We incorporate additional angular structure by adding an octopole mass moment aligned with the position angle of the power-law ellipsoid, producing boxy and disky projected density contours. We account for uncertainties in the measurements of the image positions by adding random astrometric shifts to the model image positions. The macromodel and the source position absorb stochastic astrometric perturbations associated with both measurement uncertainties and dark substructure; we can nearly always obtain a lens model (including substructure) and source coordinate that produce a lens system with the same image positions as observed in the data. 

With the model for the main deflector and dark matter halos in place, we compute flux ratios by ray-tracing through the lens system to the source plane, and integrate the flux from the extended background source. We account for uncertainties in the flux ratios by adding perturbations to model-predicted fluxes in the forward model. Next, we compute a summary statistic defined as the metric distance between the observed flux ratios, and the model predicted flux ratios
\begin{ceqn}
\begin{equation}
\label{eqn:summarystat}
    S = \sqrt{\sum_{i=1}^{3}\left(f_{\rm{data}(i)} - f_{\rm{model(i)}}\right)^2}
\end{equation}
\end{ceqn}
where $f_{\rm{model(i)}}$ is the model-predicted flux ratio, and $f_{\rm{data(i)}}$ is the observed flux ratio. To determine the posterior distribution, we employ a rejection algorithm in Approximate Bayesian Computing \citep[ABC;][]{Rubin1984}. We accept realizations if $S < \epsilon$, where $\epsilon$ is an acceptance threshold. In practice, we generate $10^5-10^6$ draws of $\bf{m}_{\rm{sub}}$ and ${\bf{x}}$ per lens, and retain the 3,000 ${\bf{q}}_{\rm{sub}}$ samples corresponding to the lowest summary statistics. This choice corresponds to $\epsilon$ values between 0.04 and 0.08. The number of accepted samples is subject to the convergence criterion that the inferred model parameters do not change when one discards random samples from the simulation and retains the same number of samples per lens (see, for example, Appendix A in \citet{Gilman2019a}). As $\epsilon \rightarrow 0$ the ratio of the number of accepted samples from the two models $\frac{N_{q1}}{N_{q2}}$ approaches the relative likelihood of the two models $\frac{\mathcal{L}\left(\bf{d}_n | \bf{q}_1\right)}{\mathcal{L}\left(\bf{d}_n | \bf{q}_2\right)}$ \citep{Rubin1984}. This method is known as an Approximate Bayesian Computing method. As we only need to know the likelihood up to a constant numerical factor to obtain the posterior distribution in Equation \ref{eqn:posterior}, this approach allows us to approximate the target posterior distribution. 

\subsection{Data}
\label{ssec:data}
The data we use in this work consists of the same set of eleven lens systems used by \citet{Gilman2022}. This sample includes nine systems with measurements of flux ratios from nuclear narrow-line emission presented by \citep{Nierenberg2014,Nierenberg2017,Nierenberg2019,Sugai07}, one system with flux ratios measured in radio wavelengths by \cite{chiba2005}, and one with CO 11-10 emission \citep{Stacey18}. Two factors determine the sample selection: First, stellar micro-lensing can impact image fluxes in wavelengths coming from physical scales around the background quasar $<0.5$ pc. The narrow-line emission, radio emission, and molecular CO emission come from spatially extended regions around the background quasar, rendering them immune to microlensing, while the size remains compact enough such that the data is sensitive to milli-lensing by dark matter halos. Second, we select lenses with main deflectors that do not show evidence of morphological complexity in the form of stellar disks, as these structures can bias the interpretation of flux ratios if the disk is not properly included in the lens model \citep{Gilman++17,Hsueh16,Hsueh17,Hsueh18}. 

\section{Modeling structure formation in ultra-light dark matter}
\label{sec:model}
Ultra-light dark matter refers to classes of dark matter theories with particle masses on the order of $10^{-22}$ eV, but the mass can vary by many orders of magnitude, up to $10^{-10}$ eV. A single parameter defines the structure formation processes associated with these theories, expressed interchangeably as either the particle mass $m_{\psi}$, or the corresponding de Broglie wavelength
\begin{ceqn}
\begin{equation}
    \lambda_{\rm{dB}} = 0.6 \left(\frac{m_\psi}{10^{-22}\,\text{eV}}\right)^{-1}
    \left(\frac{v}{200\,\text{km/s}}\right)^{-1}\,\rm{kpc}.
    \label{eq:dB}
\end{equation}
\end{ceqn}
where we interpret the velocity scale $v$ as the virial velocity of a dark matter halo. 

From Equation \ref{eq:dB}, it is evident that a particle mass of order $10^{-21}-10^{-22}$ eV will manifest quantum mechanical effects on length scales of order $1 \ \rm{kpc}$, comparable to the size of a galaxy. Simulations have borne out the implications of ULDM for cosmic structure formation \citep{Schive2014a,Schive2016,Mocz19,Li++2019,Dalal++21,Li++2021,Hui++21,Yavetz++22}. The impacts of the cosmologically-relevant de Broglie wavelength on structure formation are three-fold: First, quantum pressure acts counter to gravity, removing small-scale density fluctuations from the matter distribution of the early Universe. Thus, the matter power spectrum with ULDM exhibits a sharp cutoff below a characteristic scale. We can parameterize the suppression of small-scale structure in terms of the wavenumber $k_{1/2}$ where the transfer function from CDM to UDLM drops to one-half. This inverse length scale corresponds to a half-mode mass scale \citep{Schive2016}
\begin{ceqn}
\begin{equation}
    M_{1/2} = 3.8 \times 10^{10} \left(\frac{m_{\psi}}{10^{-22} \rm{eV}}\right)^{-4/3} M_{\odot}. \label{eq:M_half}
\end{equation}
\end{ceqn}
Similar to suppression of small-scale structure and the corresponding delay in the onset of structure formation in warm dark matter (WDM), the cutoff in the matter power spectrum suppresses both the abundance and concentrations of halos on scales comparable to and below $M_{1/2}$. Second, ULDM halos have a central soliton core, in contrast to to both CDM and WDM halos which have central density cusps. Third, ULDM halos exhibit density fluctuations throughout their volume caused by the wave-like nature of the dark matter.
\begin{figure}
    \centering
    \includegraphics[width=\linewidth]{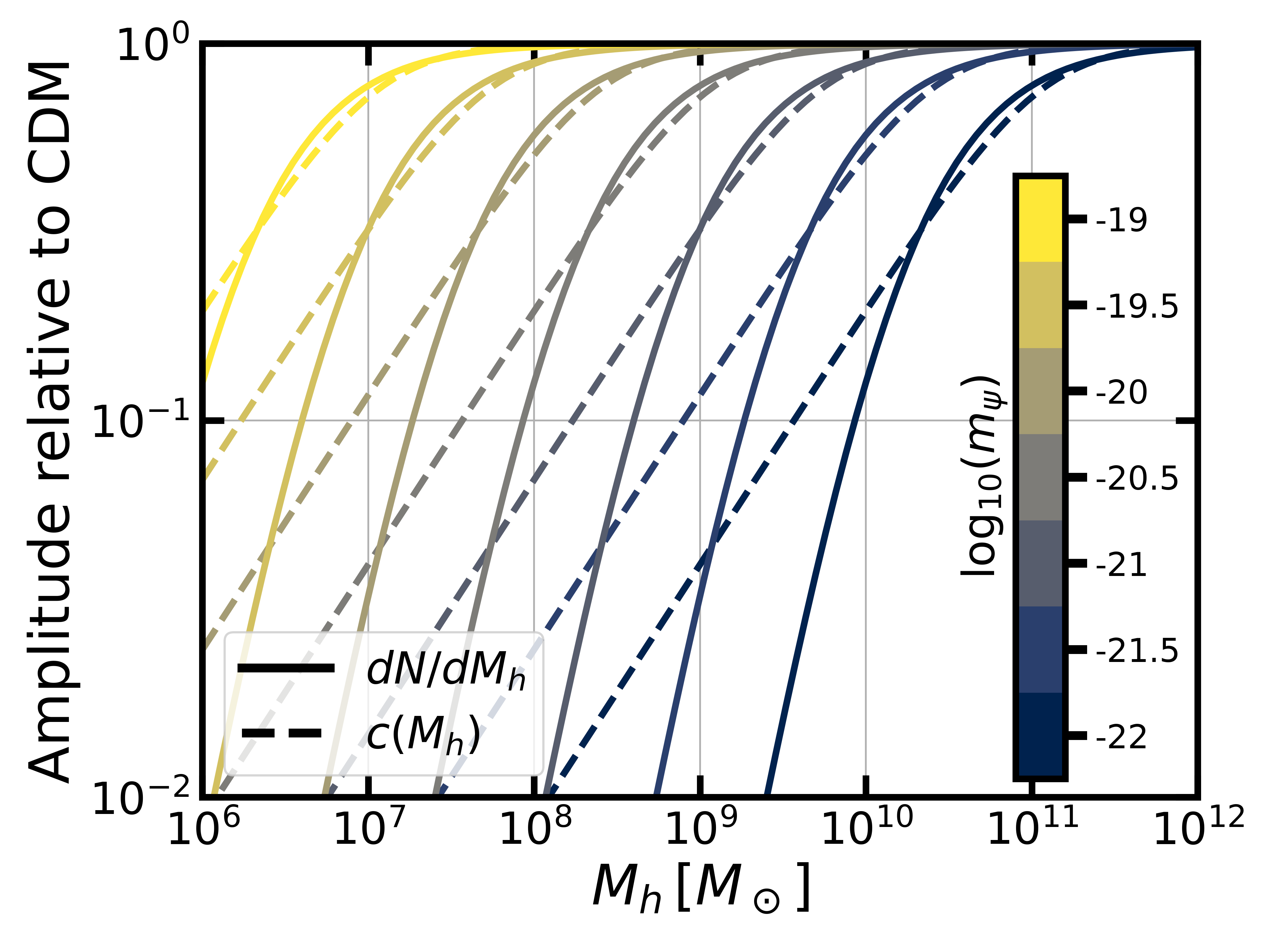}
    \caption{The subhalo mass function $dN/dM_h$ and mass-concentration relation $c\left(M_h\right)$ of ULDM, relative to CDM, for varying particle masses. Structure formation is suppressed at increasingly larger scales as $m_\psi$ decreases. Below $m_\psi \sim 10^{-18}$ eV, the properties of halos and subhalos in the mass range $10^6 - 10^{10} M_{\odot}$ we can probe with strong lensing data become practically indistinguishable.}
    \label{fig:SHMF}
\end{figure}
\begin{figure}
    \centering
    \includegraphics[width=\linewidth]{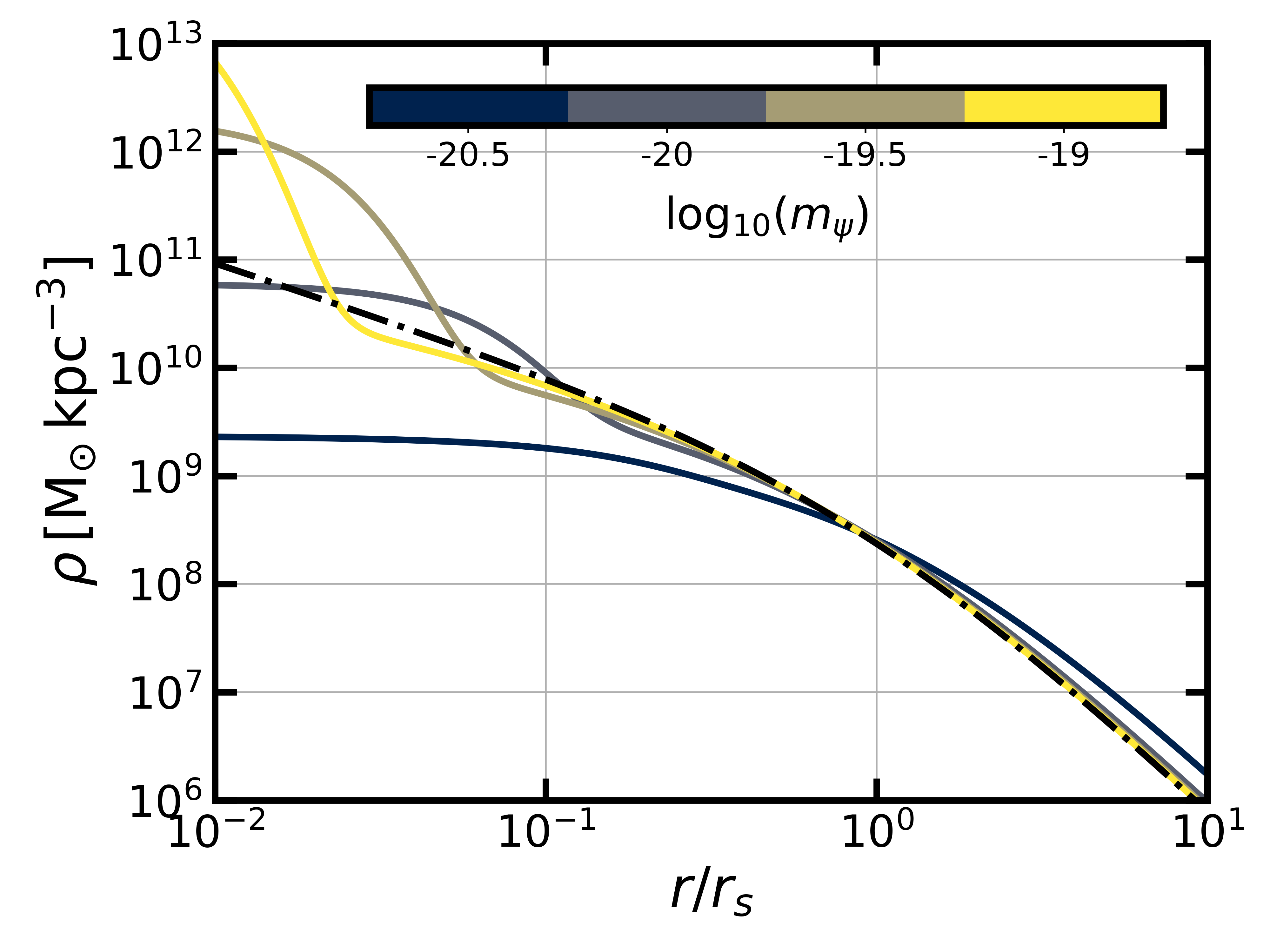}
    \caption{Density profiles of $10^8\,M_\odot$ ULDM halos for varying particle masses. A $10^8\,M_\odot$ NFW profile is plotted (black dotted-dashed line) for reference. Lighter particle masses have soliton core sizes that can exceed the scale radius $r_s$, depending on the parameter $\gamma_{\psi}$. To create the figure, we set $\gamma_{\psi} = 1/3$. The profile outside the soliton core radius $r_c$ also depends on the concentration-mass relation, which changes at fixed halo mass for varying $m_{\psi}$.}
    \label{fig:density_profiles}
\end{figure}

\begin{figure*}
    \includegraphics[trim=0.25cm 2cm 0.25cm
	2cm,width=0.48\textwidth]{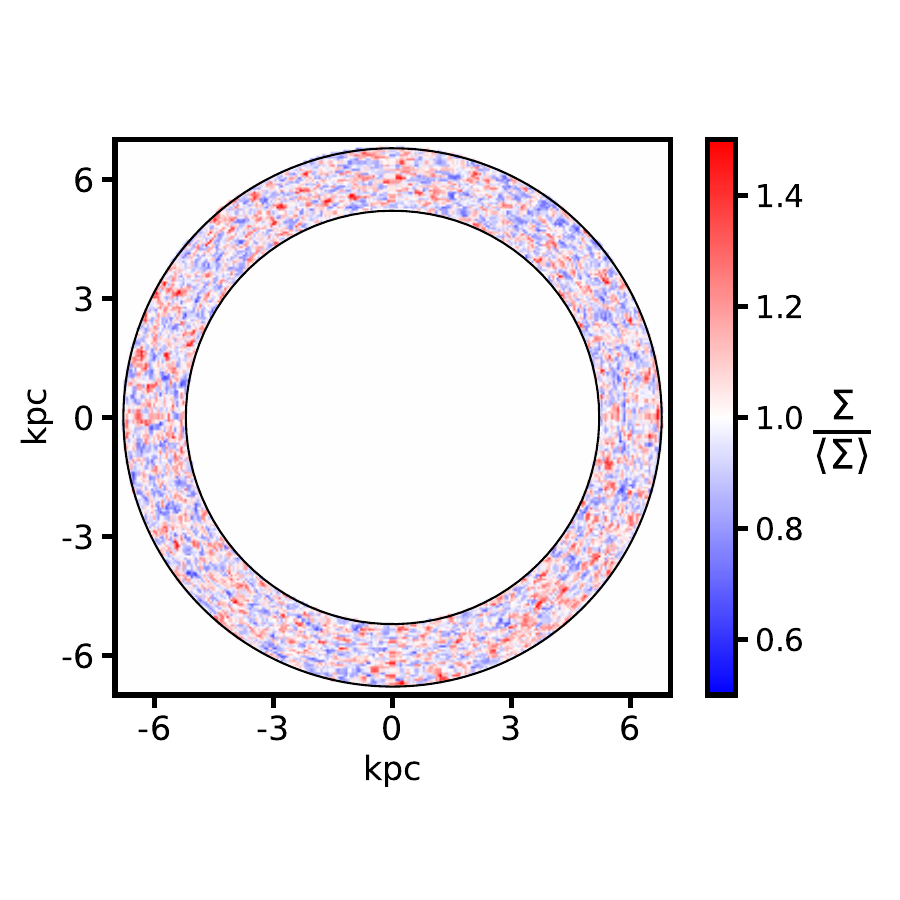}
    \includegraphics[trim=0.25cm 2cm 0.25cm
	2cm,width=0.48\textwidth]{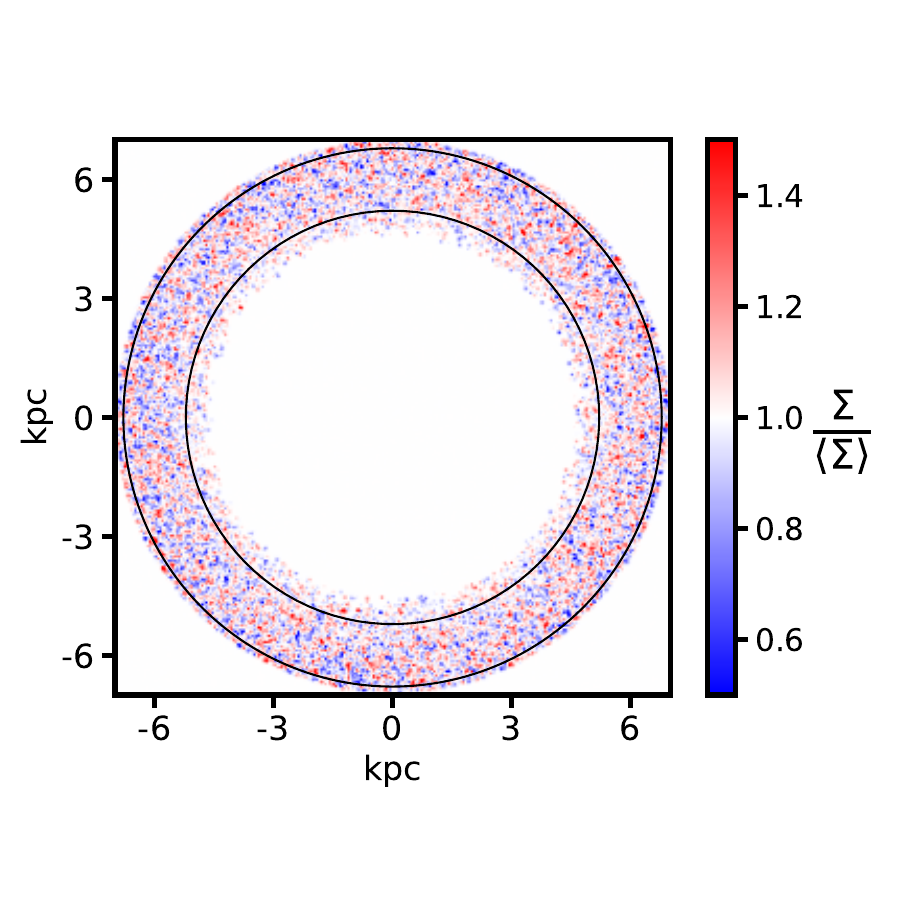}
    \caption{{\bf{Left}}: The projected mass profile of a dark matter halo simulated using the methods developed by \citet{Yavetz++22} to solve the Schr\"{o}dinger-Poisson equations, assuming a particle mass of $m_\psi = 0.8 \times 10^{-22}\,\rm{eV}$. The color scale shows variations in the projected mass of a $10^{13.3} M_{\odot}$ halo that arise due to interference effects associated with the wave-like nature of the dark matter. The circular annulus highlights the density field around $6\, \rm{kpc}$, a typical Einstein radius for a strong lens system. {\bf{Right}}: A model for the projected mass profile implemented through a superposition of circular Gaussian density profiles generated using {\tt{pyHalo}}, the software used to generate realizations of dark matter structure used in the lensing analysis.}
    \label{fig:uldmcode}
\end{figure*}

Figures \ref{fig:SHMF}, \ref{fig:density_profiles}, \ref{fig:uldmcode} and \ref{fig:proj_mass} show the halo mass function and concentration-mass relation, halo density profiles, and several two-dimensional mass maps that illustrate the changing properties of halos and fluctuations as a function of the particle mass. The following subsections give details regarding the implementation and design of this structure formation model. First, Section \ref{ssec:hmfmcrelation} details the models we implement for the (sub)halo mass function and concentration-mass relation. Second, Section \ref{ssec:profile} explains how we model the ULDM halo density profiles beneath the fluctuations. Third, Section \ref{ssec:fluctuations} details how we model the quantum fluctuations on top of the background density profile of a halo, which, as we will show, have observable consequences for flux ratio analyses with quad lenses. Fourth, Section \ref{ssec:priors} summarizes the form and motivation for the priors we implement in our analysis. Table \ref{tab:params} and \ref{tab:nuisance_params} present all of the parameters sampled in the analysis, categorized by whether they are hyper-parameters we constrain in the model (Table \ref{tab:params}) or nuisance parameters (Table \ref{tab:nuisance_params}).

\subsection{The ULDM halo mass function and concentration-mass relation}
\label{ssec:hmfmcrelation}

We model the halo and subhalo mass functions in ULDM relative to the CDM predictions. We render halos in the mass range $10^6-10^{10} M_{\odot}$. Halos less massive than $10^6 M_{\odot}$ do not affect the data due to the finite-size of the lensed background source, while halos more massive than $10^{10} M_{\odot}$ are both very rare, and very likely to host a visible galaxy, in which case we model them explicitly. 

For field halos between the observer and the source, we model the CDM mass function as
\begin{ceqn}
\begin{equation}
    \frac{dN_{\rm{CDM}}}{dM_h dV} = \delta_{\rm{LOS}} \times \xi\left(M_{\rm{host}},z\right) \frac{dN}{dM_h dV} \Big \vert_{\rm{Sheth\,Tormen}}, \label{eq:cdm_mf}
\end{equation}
\end{ceqn}
where $\delta_{\rm{LOS}}$ rescales the overall amplitude of the mass function, $\xi\left(M_h, z\right)$ adds correlated structure near the host dark matter halo \citep{Gilman2019a}, and $\frac{dN}{dM_h dV} \Big \vert_{\rm{Sheth\,Tormen}}$, is the mass function model presented by \citet{sheth2001}. Subhalos associated with the host dark matter halo around the main deflector can also impact flux ratios. We generate the population of subhalos from a mass function defined in projection

\begin{ceqn}
\begin{equation}
    \frac{dN_{\rm{CDM}}}{dM_h dA} = \frac{\Sigma_{\rm{sub}}}{M_0} \left(\frac{M_h}{M_0}\right)^{-\alpha} \mathcal{F}\left(M_{\rm{host}},z\right),
\end{equation}
\end{ceqn}
where $\Sigma_{\rm{sub}}$ (with dimension $\rm{kpc^{-2}}$) sets the amplitude at the pivot scale $M_0 = 10^8 M_{\odot}$, $\alpha$ is the logarithmic slope of the mass function, and $\mathcal{F}\left(M_{\rm{host}},z\right)$ accounts for the evolution in the projected mass density of subhalo with host halo mass and redshift \citep{Gilman2019a}. By factoring out the dependence on host halo mass and redshift, we can interpret the normalization $\Sigma_{\rm{sub}}$ as a hyper-parameter predicted by $\Lambda$CDM, and combine inferences of it from mutiple lenses.
\begin{table*}
		\centering
		\caption{Hyper-parameters sampled in the forward model. Notation $\mathcal{U} \left(u_1, u_2\right)$ indicates a uniform prior between $u_1$ and $u_2$. These hyper-parameters are referred to as ${\bf{q}}_{\rm{sub}}$ in Equations \ref{eqn:posterior} and \ref{eqn:likelihood}. We obtain their joint distribution by multiplying individual likelihoods from each lens.}
		\label{tab:params}
		\begin{tabular}{lccr} % four columns, alignment for each
			\hline
			parameter & definition & prior\\
			\hline 
			$\log_{10}(m_\psi/\rm{eV})$ & ultra-light boson mass
			(Equation \ref{eq:dB})& $\mathcal{U}(-22.5,-16.5)$ 
            \\\\
            $\gamma_\psi$ & power law exponent for core radius-halo mass relation (Equation \ref{eq:r_c})& $\mathcal{U}(0.2,0.5)$
            \\\\
            $\log_{10}(A_{\rm{fluc}})$ & fluctuation amplitude 
            (Section \ref{ssec:fluctuations}) & $\mathcal{U}(-3.5,-0.5)$ \\
			\\
			$\Sigma_{\rm{sub}} \left[\rm{kpc}^{-2}\right]$ & normalization of subhalo mass function (Equation \ref{eq:cdm_mf})&  $\mathcal{U} \left(0, 0.1\right)$ \\&(rendered between $10^6-10^{10} \rm{M}_\odot$) & \\
			\\
			$\alpha$ & logarithmic slope of the subhalo mass function & $\mathcal{U} \left(-1.95, -1.85\right)$\\
			\\
			\\
			$\delta_{\rm{los}}$ & rescaling factor for the line of sight Sheth-Tormen & $\mathcal{U} \left(0.8, 1.2\right)$ \\
			&mass function (Equation \ref{eq:cdm_mf}, rendered between $10^6-10^{10} \rm{M}_\odot$)&\\
			\hline		
		\end{tabular}
\end{table*}

\begin{table*}
		\centering
		\caption{Nuisance parameters sampled in the forward model. These hyper-parameters are referred to as ${\bf{x}}$ in Equation \ref{eqn:likelihood}. We marginalize over these parameters to compute the likelihood of a single lens' data before multiplying the likelihoods to compute the posterior distribution. Lens-specific priors are summarized in Table 2 of \citet{Gilman2019a}.}
		\label{tab:nuisance_params}
		\begin{tabular}{lccr} % four columns, alignment for each
			\hline
			parameter & definition & prior\\
			\hline 
			$\log_{10} \left(M_{\rm{host}}/\rm{M}_\odot\right)$ & host halo mass &  (lens specific) \\
			\\
			$s$ & size of an individual fluctuation relative to $\lambda_{\rm{dB}}$ & $\mathcal{U}\left(0.025,  0.075\right)$ \\
			\\
			$a_4$ & octopole moment of main deflector mass profile & $\mathcal{N}\left(0.0, 0.01\right)$ \\ 
			& (introduces boxyness and diskyness) & \\
			\\
			$\sigma_{\rm{src}} \left[\rm{pc}\right]$ & source size, full-width at half maximum of a Gaussian& \\
			& narrow-line & $\mathcal{U} \left(25, 60\right)$\\
			& radio & $\mathcal{U} \left(1, 5\right)$ \\
			& CO 11-10 & $\mathcal{U} \left(1, 20\right)$ \\
			\\
			$\gamma_{\rm{macro}}$ & logarithmic slope of main deflector mass model  & $\mathcal{U} \left(1.95, 2.2\right)$ \\
			\\
			$\gamma_{\rm{ext}}$ & external shear strength in the main lens plane & (lens specific) \\
			\\
			$\delta_{xy} \left[\rm{m.a.s.}\right]$ & image position uncertainties& (lens specific)\\
			\\
			$\delta f$ & image flux or flux ratio uncertainties & (lens specific)\\
			\hline		
		\end{tabular}
\end{table*}

In CDM, structure formation is scale-free over many orders of magnitude in halo mass, with a minimum halo mass comparable to the mass of the Earth \citep{green2005}. In ULDM, quantum pressure associated with the large de Broglie wavelength of ULDM particles precludes the formation of low mass halos, introducing a minimum halo mass scale \citep{Schive2014b}

\begin{ceqn}
\begin{eqnarray}
\label{eq:M_min}
    \nonumber M_{\rm{min}}(z) &= 1.2\times10^8 \left(\frac{m_{\psi}}{10^{-22}\rm{eV}}\right)^{-3/2} \\ &\times\left(1+z\right)^{3/4}
    \left(\frac{\zeta(z)}{\zeta(0)}\right)^{1/4}\,M_\odot,
\end{eqnarray}
\end{ceqn}
where $\zeta(z)\equiv(18\pi^2+82(\Omega_m(z)-1)-39(\Omega_m(z)-1)^2)/\Omega_m(z),$ and $\Omega(z)$ is the matter density parameter. When rendering populations of ULDM halos, we add a sharp cutoff in the mass function at $M_{\rm{min}}\left(z\right)$, rendering no halos with mass below this scale. 

For the remaining population of halos more massive than $M_{\rm{min}}$, the abundance relative to CDM is suppressed by a factor 
\begin{ceqn}
\begin{equation}
\label{eqn:suppression}
    F\left(x\right) = \left[1+a x^b\right]^c,
\end{equation}
\end{ceqn}
such that the mass function in ULDM can be parameterized relative to CDM through the relation
\begin{ceqn}
\begin{equation}
\label{eqn:mfuncsuppression}
    \left.\frac{dn}{dM_h}\right|_{ULDM} =
    \left.\frac{dn}{dM_h}\right|_{CDM} \times
    F\left(\frac{M_h}{M_{1/2}}\right),
\end{equation}
\end{ceqn}
where $(a, b,c)=(0.36, -1.1,-2.2)$ \citep{Schive2016}\footnote{The prefactor $a=0.36$ comes from converting the mass scale $M_{\psi}$ presented by \citet{Schive2016} to $M_{1/2}$ using their Equations 6 and 7.}. This parameterization is similar to a parameterization of the mass function in WDM models, although the physical mechanisms responsible for the suppression differ. Figure \ref{fig:SHMF} shows
the suppression of the halo mass function relative to CDM as a function of the halo mass for different values of $m_\psi.$ We apply the same suppression term (Equation \ref{eqn:suppression}) to populations of ULDM subhalos and field halos. 

Tidal stripping complicates the modeling of the subhalo mass function, because tidal interactions between subhalos and the host halo can both remove mass from subhalos, effectively destroying them \citep[e.g.][]{Fiacconi++16,Garrison-Kimmel++17,WebbBovy20}, and deform their density profiles \citep[e.g.][]{Errani2019,green2019,Du2018}. One arguably conservative way to phrase the complicated effects associated with tidal stripping of ULDM subhalos is that we simply do not know the amplitude of the subhalo mass function. Thus, we account for the effects of tidal stripping by marginalizing our results over $\Sigma_{\rm{sub}}$ in the range $0.0 - 0.1 \ \rm{kpc^{-2}}$\footnote{For reference, \citet{Nadler++21} showed that the $\Lambda$CDM prediction for the subhalo mass function amplitude is $0.025 \  \rm{kpc^{-2}}$ ($0.05 \ \rm{kpc^{-2}}$) if tidal stripping is equally efficient (doubly-efficient) in the Milky Way, relative to the elliptical galaxies that tend to act as strong lenses.}. 

By eliminating low-mass halos, the suppression of small-scale power in ULDM models breaks the hierarchical structure formation process that characterizes CDM. In addition to altering the (sub)halo mass function, this also delays the onset of structure formation and the buildup of halo mass through accretion of smaller structures. As the central density of a halo reflects the background density of the Universe when the constituents formed \citep{Bullock_2001,Wechsler_2002}, the delayed onset of structure formation therefore suppresses the concentration-mass relation, relative to the CDM prediction \citep{Schneider++15}. 

The dashed lines in Figure \ref{fig:SHMF} show the concentration-mass relations for $m_\psi\in[10^{-22},10^{-19}]\,\text{eV}$ relative to the CDM prediction. We computed the ULDM concentration-mass relation using $\tt{galacticus}$ \citep{Benson12}, applying the algorithm presented by \citet{Schneider++15} to ULDM transfer functions. Specifically, the collapse time of a ULDM halo is computed from extended Press–Schechter theory \citep{Press:1973iz,Bond:1990iw}. The halo is then assigned a concentration that a CDM halo with the same collapse time would have. Relative to the transfer functions relevant for warm dark matter \citep[e.g.][]{Schneider++15,Bose++16}, the resulting concentration-mass relations in ULDM have a sharper cutoff, and turnover happens closer to the mass scale where the halo mass function becomes suppressed. These effects can be traced back to the shape of the ULDM transfer function, which has a sharper cutoff than the WDM transfer functions computed for thermal relic or sterile neutrino dark matter \citep{Abazajian++19}. We model the concentration-mass relation in ULDM by multiplying the CDM concentration-mass relation by Equation \ref{eqn:suppression},
\begin{ceqn}
\begin{equation}
   c_{\rm{ULDM}}(M,z) = c_{\rm{CDM}}(M,z) \times
   F\left(\frac{M_h}{M_{1/2}}\right),
    \label{eq:mcrel}
\end{equation}
\end{ceqn}
where in this case the cutoff function $F\left(x\right)$ has $(a, b, c)=\left(3.348, -0.489, -1.5460\right)$. To implement the CDM concentration-mass relation, we use the model presented by \citet{DiemerJoyce19} with a scatter of 0.2 dex. We note that, by applying the \citet{Schneider++15} algorithm to compute the ULDM concentration-mass relation, we assume that the quantum-mechanical effects that produce the soliton core and cause fluctuations in the host halo do not alter concentration-mass relation, which affects the halo density profile outside the soliton core. Put differently, we assume the concentration-mass relation depends only on the transfer function that alters the shape of the linear matter power spectrum. Although this assumption has not yet been verified with numerical simulations, it is fairly well-motivated by the fact that the concentration-mass relation depends primarily on the accretion history of a halo and its formation time \citep[e.g.][]{Bullock_2001,Wechsler_2002,DiemerJoyce19,Wang++20}, and these processes transpire on much longer timescales than $\lambda_{\rm{dB}} v^{-1} \sim 10^{-3} \left(\frac{10^{-22} \rm{eV}}{m_{\psi}}\right) \left(\frac{200 \ \rm{km} \ \rm{s^{-1}}}{v}\right)^2$ Gyr, the relevant timescale for the quantum fluctuations of the host halo density profile\footnote{The timescale for fluctuations is much less than the light crossing time of the halo for $m_\psi>10^{-21}\,\rm{eV},$ and could potentially affect image light curves. An interesting avenue for future work would be to investigate if the fluctuation evolution timescale can affect time delays between lensed images.}.

\begin{figure}
    \includegraphics[width=0.45\textwidth]{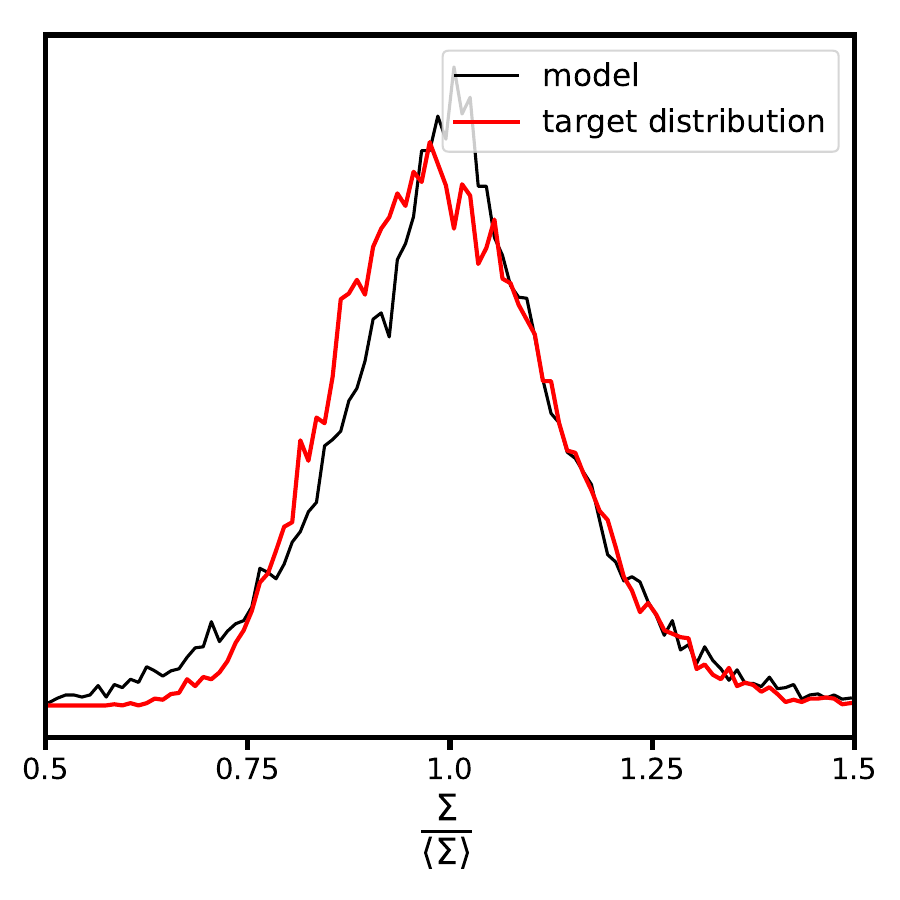}
    \caption{The distribution of $\Sigma / \langle \Sigma \rangle$, the projected mass density normalized by the mean projected mass density at the center of the circular annulus shown in Figure \ref{fig:uldmcode}, for the two mass distributions shown in Figure \ref{fig:uldmcode}. The red and black curves show the distribution of $\Sigma / \langle \Sigma \rangle$ for the numerically-computed projected mass distribution (left panel of Figure \ref{fig:uldmcode}), and the model realization (right panel), respectively.}
    \label{fig:uldmcodepdfs}
\end{figure}

\subsection{ULDM halo density profile}\label{ssec:profile}
Flux ratios are sensitive to the internal structure of dark matter halos \citep[e.g.][]{Nierenberg2014,Gilman2019a,Gilman2021}. In contrast to the Navarro-Frenk-White (NFW) \citep{Navarro++97} halos predicted by CDM, which have cuspy central density profiles, ULDM halos have flat central profiles, known as soliton cores. The soliton core radius is related to the particle mass by \citep{Schive2014b}
\begin{ceqn}
\begin{equation}
    r_c = 1.6\left(\frac{m_{\psi}}{10^{-22}\rm{eV}}\right)^{-1}
    \frac{1}{(1+z)^{1/2}}\left(\frac{\zeta(z)}{\zeta(0)}\right)^{-1/6}\left(\frac{M_h}{10^9\,M_\odot}\right)^{-\gamma_{\psi}}\,\text{kpc},
    \label{eq:r_c}
\end{equation}
\end{ceqn}
where $\gamma_{\psi}$ is the power law exponent of the core radius-halo mass relation, typically $\gamma_{\psi}\sim0.2 - 0.5.$ Different simulations yield different core-halo relations \citep[e.g.][]{Chan22,Schive2014b,Mocz19,Mina20,Schwabe16,Nori20,Yavetz++22,Glennon++22}. For a summary of core-halo structures found in the literature, see Section 4.2.1 of \citet{Chan22}. We note that \citet{DeLaurentis_Salucci22}, motivated by \citet{Burkert20}, determined that the DM halo density core of the M87 galaxy is inconsistent with Equation \eqref{eq:r_c}.

\citet{Schive2014a} present a fitting function for the density profile of the soliton
\begin{ceqn}
\label{eqn:soliton}
\begin{equation}
    \rho_c(x)=\frac{1.9a^{-1} \left(m_{\psi} / 10^{-23}\,\rm{eV}\right)^{-2}(x_c/\text{kpc})^{-4}}{[1+9.1\times10^{-2}(x/x_c)^2]^8}\,M_\odot/\text{pc}^{3}, 
    \label{eq:rho_c}
\end{equation}
\end{ceqn}
where $x=r/a,$ $x_c=r_c/a$ and $a = \left(1+z\right)^{-1}$. At radii $r \gg r_c$, the ULDM halo density profile transitions from the soliton profile to an NFW profile. We replicate this transition through a superposition of the soliton profile with a cored NFW profile
\begin{ceqn}
\begin{equation}\label{eq:rho_n}
    \rho_n(r, \tilde{r}_c) = \frac{\rho_s r_s^3}{(r+\tilde{r}_c)(r+r_s)^2}, 
\end{equation}
\end{ceqn}
where $\rho_s$ and $r_s$ are the density normalization and scale radius of an NFW profile, and $\tilde{r}_c$ is a core radius for the NFW halo. 

We construct the halo density profile by superimposing the density profiles in Equations \eqref{eq:rho_c} and \eqref{eq:rho_n} subject to the constraints
\begin{ceqn}
\begin{eqnarray}
    \rho_c(0) &=& \rho_n(0) + q\rho_c(0)\\
     M_{200} &=& M_n(r<r_{200}) + qM_c(r<r_{200}),
\end{eqnarray}
\end{ceqn}
which enforces that the central density match the central soliton density predicted by Equation \ref{eqn:soliton}, and mass conservation within $r_{200}$, respectively. Solving for the $\tilde{r}_c$ and $q$ that satisfy these constraints gives a density profile
\begin{ceqn}
\begin{equation}
    \rho(r)=\rho_n(r, \tilde{r}_c) + q \rho_c(r).
    \label{eq:profile}
\end{equation}
\end{ceqn}
Resulting density profiles for a $10^8\,M_\odot$ halo for $m_\psi\in[10^{-20.5},10^{-19}]\,\text{eV}$ and $\gamma_{\psi} = 1/3$ are shown in Figure \ref{fig:density_profiles}. In our analysis, we marginalize over $\gamma_{\psi}$ to account for uncertainties in the connection between the core mass and the total halo mass (see Section \ref{ssec:priors}).

\subsection{Modeling quantum fluctuations in the dark matter profile of the host halo}\label{ssec:fluctuations}
Due to the kpc-scale de Broglie wavelength, ULDM halos exhibit quantum density fluctuations throughout their volume \citep{Church++19,Schive2014a,Hui2021,kawai2022,Dalal++22,Schwabe++21,Chowdhury++21,Lancaster++20}. \citet{Chan2020} pointed out that these fluctuations should impart measurable perturbations to the flux ratios in quadruply-imaged quasars. If this is the case, then omitting fluctuations from the lens model could cause one to conflate signal from the density fluctuations with perturbations by dark matter halos, biasing inferences of the parameters that define the mass function and concentration-mass relation, as well as the particle mass itself.

The left panel of Figure \ref{fig:uldmcode} shows density fluctuations isolated near $6 \ \rm{kpc}$, a typical Einstein radius for a strong lens system, for a $10^{13.3} M_{\odot}$ halo simulated with a particle mass of $0.8 \times 10^{-22}\,\rm{eV}$ using the methods presented by \citet{Yavetz++22} to solve the Schr\"{o}dinger-Poisson equations. Our goal is to develop a model from which we can generate many realizations of projected density fields that share the same statistical properties as the projected density field shown in the left panel. 

To begin, we comment on several general features of the fluctuation density field that motivate certain aspects of our model. First, the size of an individual fluctuation should scale proportionally with $\lambda_{\rm{dB}}$, and therefore inversely with the particle mass, although the size of a fluctuation is not necessarily close to or equal to $\lambda_{\rm{dB}}$ as the structure of an individual patch of projected mass density includes many individual fluctuations distributed along the line of sight through the halo. Second, the expected amplitude of a fluctuation about $\langle \Sigma \rangle$, the mean projected density of the host halo at the Einstein radius, should scale as $\sqrt{\lambda_{\rm{dB}}}$. This follows from the central limit theorem. When many individual fluctuations are viewed in projection, their contributions to the net projected density are random variables, and thus a measurement of $\delta \Sigma \equiv \Sigma / \langle \Sigma \rangle$ at any random position samples a Gaussian distribution with mean zero and a (root) variance that scales as $\frac{1}{\sqrt{N}}$, where $N$ is the number of fluctuations along the line of sight projected through the host halo. The number of fluctuations along the line of sight through the host halo scales as $\frac{1}{\lambda_{\rm{dB}}}$, such that the variance of the density field scales as $\sqrt{\lambda_{\rm{dB}}}$. Analytic predictions for $\delta \Sigma$ presented by \citet{Chan2020} and \cite{kawai2022} also exhibit the $\sqrt{\lambda_{\rm{dB}}}$ scaling of the variance of the density field. 

Knowing how the size and amplitude of the fluctuations should scale with the particle mass, we implement a hierarchical framework to describe the fluctuation density field. We model individual fluctuations as circular Gaussian profiles. The amplitudes of the individual profiles are drawn from a Gaussian distribution with mean zero and a variance $A_{\rm{fluc}} \sqrt{\lambda_{\rm{dB}}}$, where $A_{\rm{fluc}}$ is a hyper-parameter in our model. The size of each individual fluctuation is drawn from another Gaussian distribution with mean $s \lambda_{\rm{dB}}$ and variance $s^{\prime} \lambda_{\rm{dB}}$. In defining the hyper-parameters $A_{\rm{fluc}}, s$, and $s^{\prime}$, we have enforced the correct scaling with the de Broglie wavelength discussed in the previous paragraph. Finally, we distribute these profiles randomly in two dimensions with a number density per unit area $s^{-2} \lambda_{\rm{dB}}^{-2}$. When analyzing the data, we generate the fluctuations in circular apertures with radii of 0.2 arcsec around each lensed image.

To determine appropriate values of $A_{\rm{fluc}}, s, \ \rm{and} \ s^{\prime}$, we compare density fields simulated from the model with an exact realization of the density field of a $10^{13.3} M_{\odot}$ host halo. In particular, we calibrate our model to reproduce the distribution of $\Sigma / \langle \Sigma \rangle$ because fluctuations in the projected mass near an image cause flux ratio perturbations in strong lenses. The model realization shown in the right panel of Figure \ref{fig:uldmcode}, generated with $\log_{10} (A_{\rm{fluc}}) = -1.3$, $s = 0.05$, and $s^{\prime} = 0.2$, results in a distribution of $\Sigma / \langle \Sigma \rangle$, shown in Figure \ref{fig:uldmcodepdfs}, that matches the numerical result to approximately $7 \%$. The distribution of $\Sigma / \langle \Sigma \rangle$ is relatively insensitive to the parameter $s^{\prime}$ provided $s^{\prime} \approx s$, so we fix $s^{\prime} = 0.2$. Independently, \citet{Chan2020} arrived at a prediction, based on the simulations of \citet{Schive2014b}, for the fluctuation amplitude $\log_{10} (A_{\rm{fluc}}) = -1.3$, in good agreement with our numerical calculation.  \citet{kawai2022} predict a lower amplitude of the fluctuations, corresponding to $\log_{10}( A_{\rm{fluc}} )= -1.6$\footnote{To compute this number, we used Equation 30 in \citet{kawai2022} with a circular top-hat window function to compute the variance.}.

Finally, to compute the amplitude of the fluctuations in the total projected mass including baryons, we multiply $A_{\rm{fluc}}$ by $f$, where $f$ is the fraction of dark matter to baryonic matter at the Einstein radius. By analyzing a sample of 21 strong lensing elliptical galaxies, \citet{Shajib++21} find $f \approx 0.48 \pm 0.15$\footnote{This result was obtained through private communication with the lead author.}. The presence of baryonic matter effectively suppresses the amplitude of the fluctuations in the total projected mass. Combining the theoretical prediction for $A_{\rm{fluc}}$ with the contribution from baryons, we have the theoretical expectation for the fluctuation amplitude $\log_{10}(A_{\rm{fluc}}) = -1.6$. The uncertainty in $f$ contributes 0.1 dex uncertainty in this value, while we estimate from numerical calculations an additional 0.15 dex scatter between realizations of the density field of $\Sigma / \langle \Sigma \rangle$, so we set the uncertainty in $\log_{10}(A_{\rm{fluc}})$ to 0.2 dex. 

We now generalize this model to lenses with different host halo masses at different redshifts. We will express the result in units of convergence, or projected mass normalized by $\Sigma_{\rm{crit}}$, the critical surface mass density for lensing. The simulated halo used to match the numerical calculation shown in Figure \ref{fig:uldmcode} has a projected mass density in dark matter $\Sigma_{\rm{host}} = 8.0 \times 10^{8} M_{\odot} \rm{kpc^{-2}}$, with a lens (source) redshift of 0.5 (1.5), giving $\Sigma_{\rm{crit}} = 2.3 \times 10^9 M_{\odot} \rm{kpc^{-2}}$. The expected standard deviation of fluctuations in the convergence, $\sqrt{\langle \delta \kappa^2 \rangle}$, is given by 
\begin{eqnarray}
\label{eqn:flucamp}
    \nonumber \sqrt{\langle \delta \kappa^2 \rangle} &=& 0.025 \left(\frac{A_{\rm{fluc}}}{0.05}\right) \left(\frac{f}{0.5}\right) \left(\frac{m_{\psi}}{10^{-22} \rm{eV}} \right)^{-1/2} \\
    && \times \left(\frac{\Sigma_{\rm{host}}}{8.0\times10^8 M_{\odot}\rm{kpc^{-2}}}\right) \left(\frac{2.3 \times 10^{9} M_{\odot} \rm{kpc^{-2}}}{\Sigma_{\rm{crit}}}\right).
\end{eqnarray}
Since the amplitude of the fluctuations scales proportionally with $\Sigma_{\rm{host}}/\Sigma_{\rm{crit}}$, we can safely ignore fluctuations in the mass profile of subhalos and field halos, as $\Sigma_{\rm{host}}/\Sigma_{\rm{crit}}\ll1$ for these objects.

Strong lenses are characterized by super-critical densities, i.e. $\kappa > 1$, so we expect these fluctuations will not produce multiple images. However, their amplitudes are large enough that we expect them to significantly affect the flux ratios. To understand why, we can consider the radially-averaged central density inside the scale radius, a quantity we associate with lensing efficiency, of a typical $10^8 M_{\odot}$ halo at $z = 0.5$. Computing this density for a halo with $\rho_s = 2.4\times10^7 M_{\odot} \ \rm{kpc^{-3}}$ and $r_s = 0.57 \ \rm{kpc}$, and dividing by $\Sigma_{\rm{crit}}$, we have a fluctuation in the convergence from the halo of 0.007, approximately one-third the amplitude of a typical fluctuation in the host halo density profile associated with ULDM. Thus, we expect the fluctuations to cause perturbation to flux ratios that matches or exceeds the contribution to the signal from $10^8 M_{\odot}$ halos.
\begin{figure*}
    \centering
    \begin{subfigure}{0.33\linewidth}
        \centering
        \includegraphics[width=\linewidth]{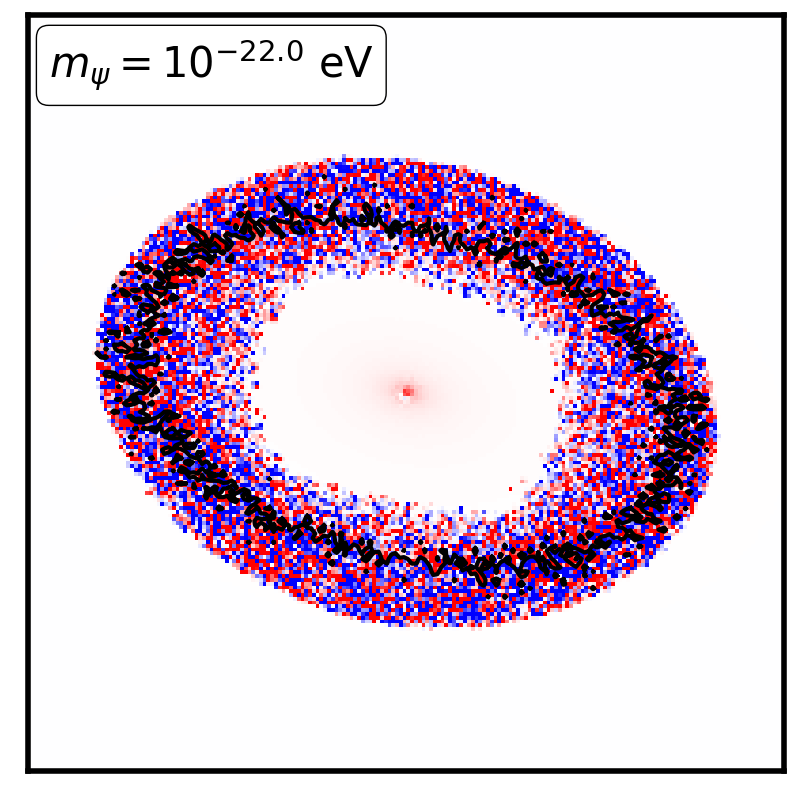}
    \end{subfigure}
    \hfill
    \begin{subfigure}{0.33\linewidth}
        \centering
        \includegraphics[width=\linewidth]{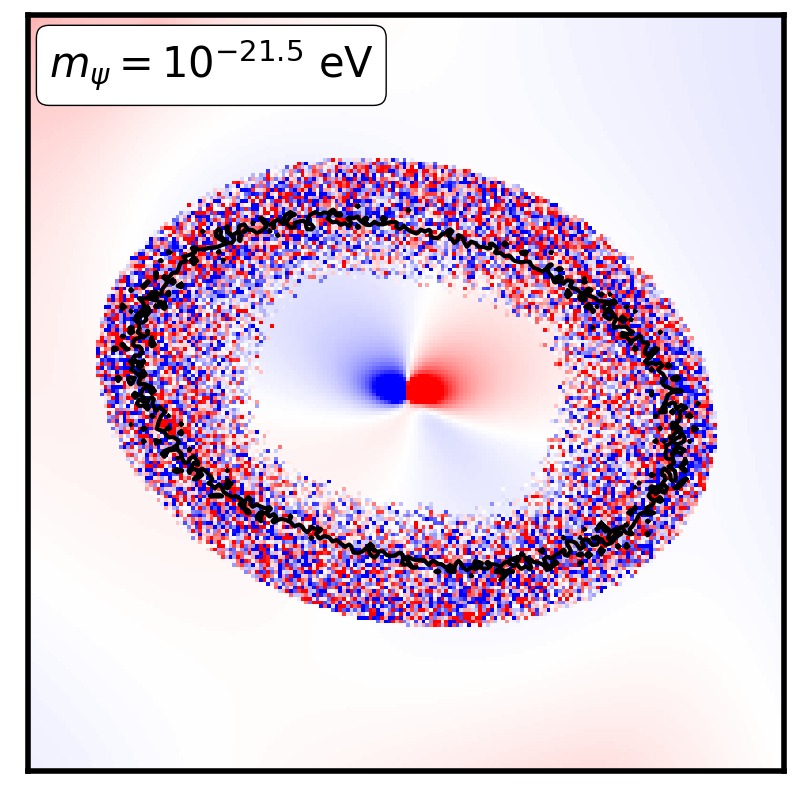}
    \end{subfigure}
    \hfill
    \begin{subfigure}{0.33\linewidth}
        \centering
        \includegraphics[width=\linewidth]{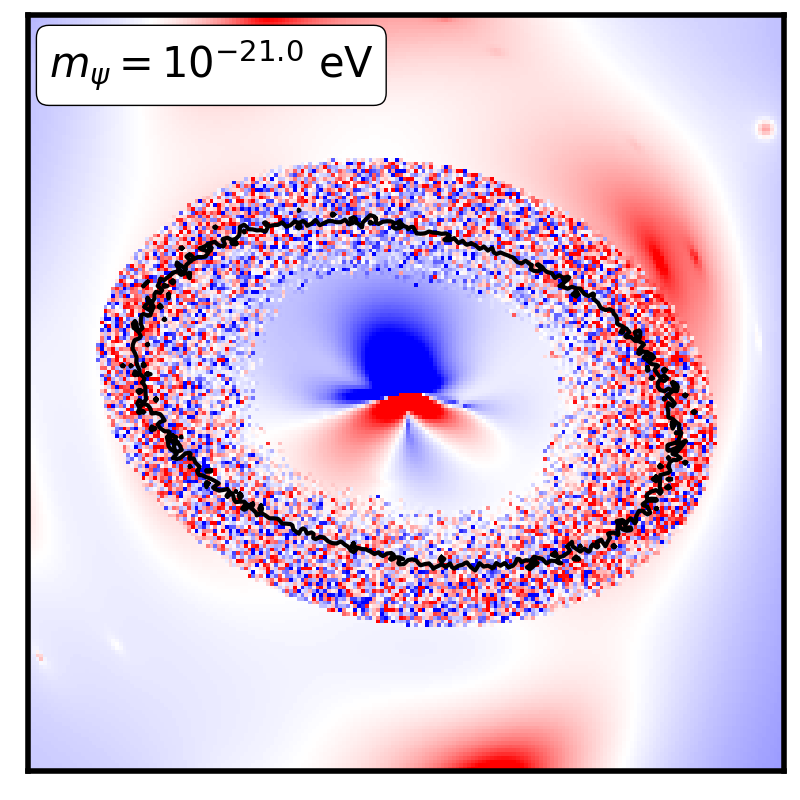}
    \end{subfigure}
    \centering
    \begin{subfigure}{0.33\linewidth}
        \centering
        \includegraphics[width=\linewidth]{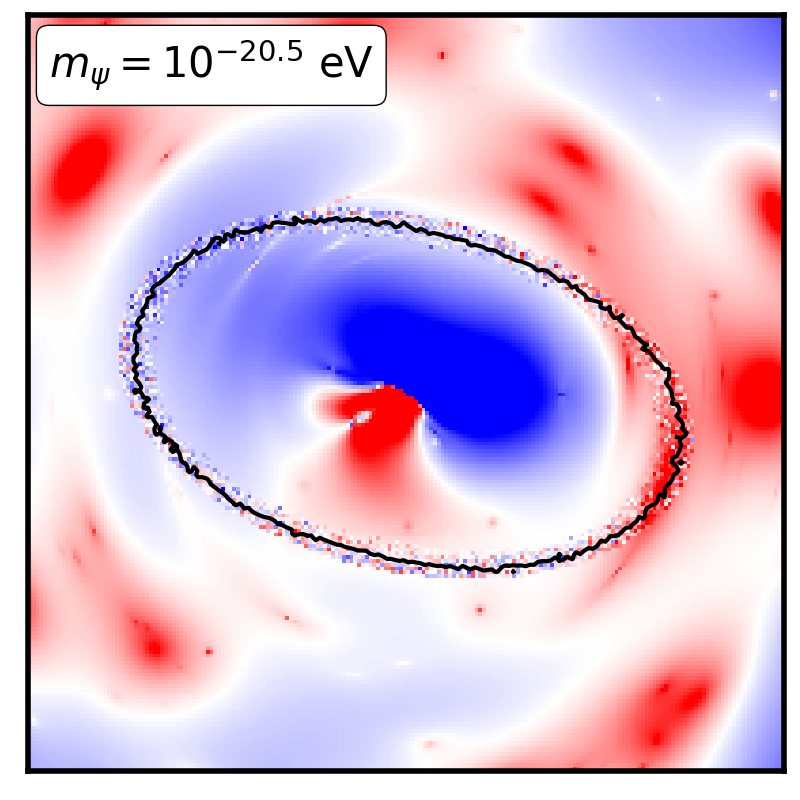}
    \end{subfigure}
    \hfill
    \begin{subfigure}{0.33\linewidth}
        \centering
        \includegraphics[width=\linewidth]{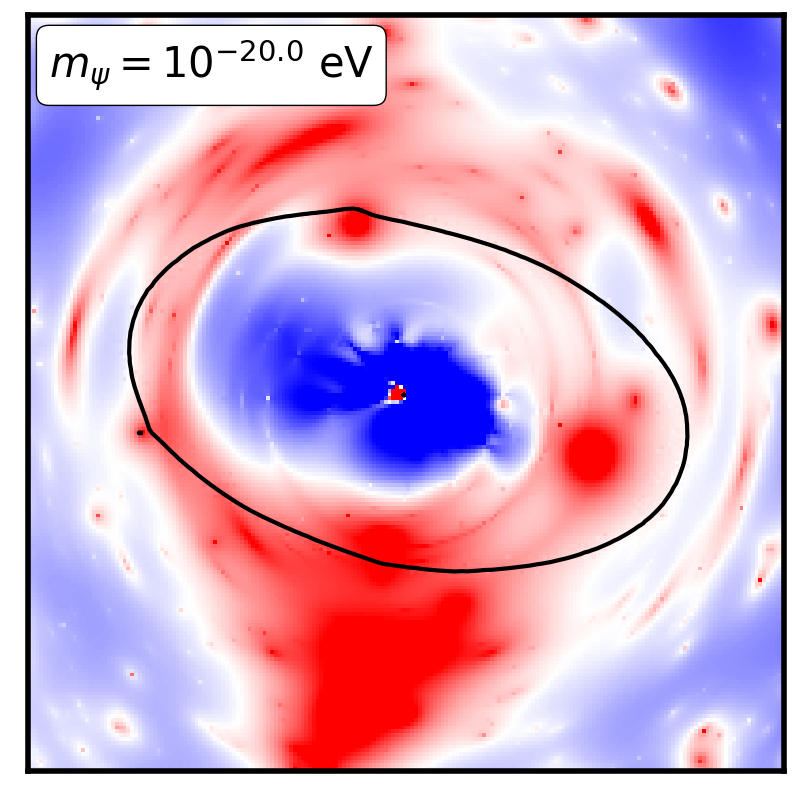}
    \end{subfigure}
    \hfill
    \begin{subfigure}{0.33\linewidth}
        \centering
        \includegraphics[width=\linewidth]{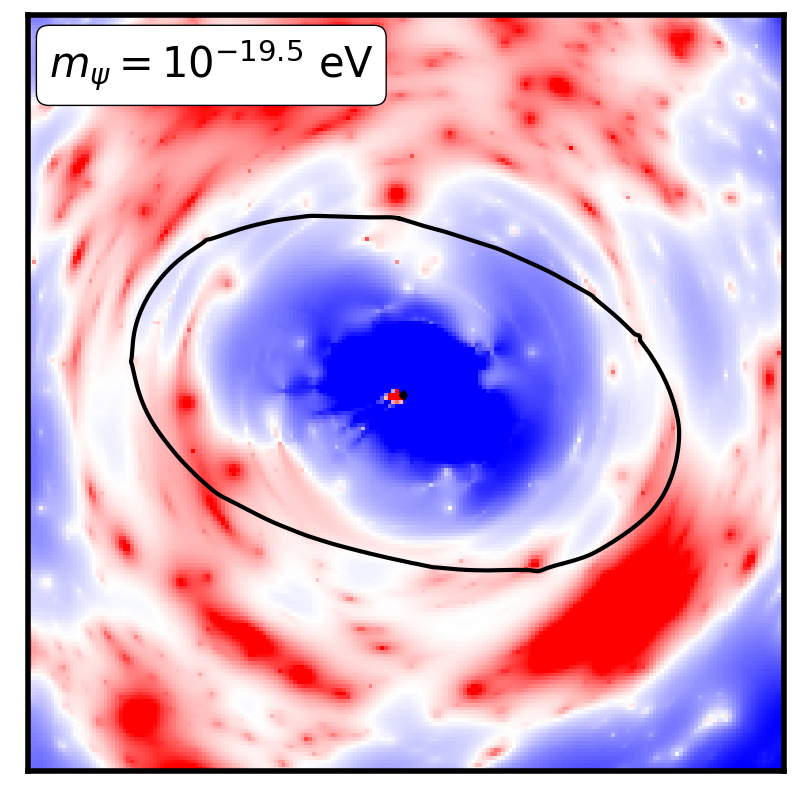}
    \end{subfigure}
    \centering
    \begin{subfigure}{0.33\linewidth}
        \centering
        \includegraphics[width=\linewidth]{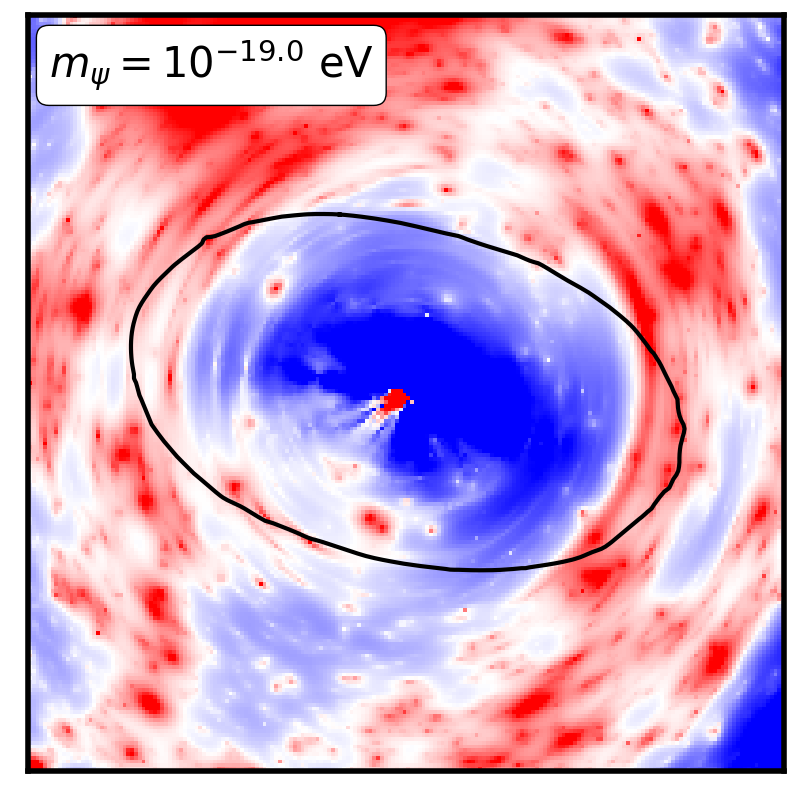}
    \end{subfigure}
    \hfill
    \begin{subfigure}{0.33\linewidth}
        \centering
        \includegraphics[width=\linewidth]{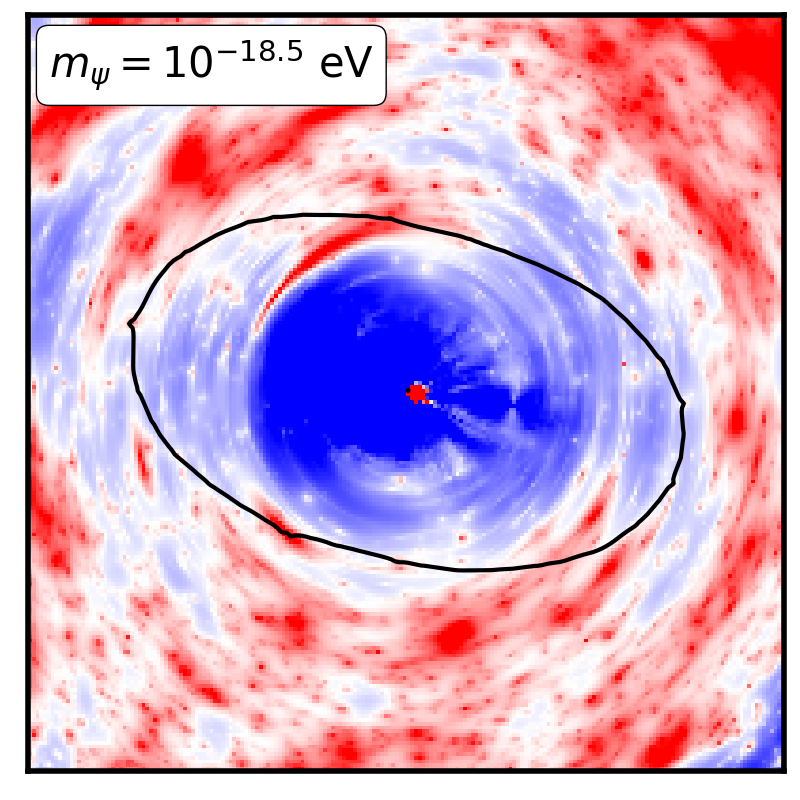}
    \end{subfigure}
    \hfill
    \begin{subfigure}{0.33\linewidth}
        \centering
        \includegraphics[width=\linewidth]{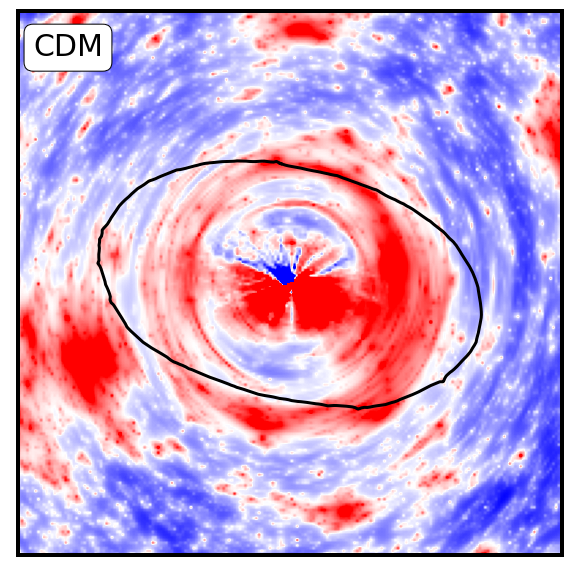}
    \end{subfigure}
    \centering
    \begin{subfigure}{0.35\linewidth}
        \centering
        \includegraphics[width=\linewidth]{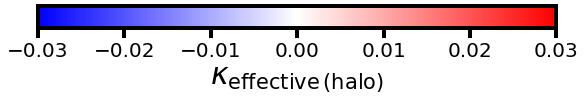}
    \end{subfigure}
    \caption{Dark matter halo \textit{effective multi-plane convergence} maps of ULDM structure for varying $m_\psi$, in comparison to CDM (bottom right). The \textit{effective multi-plane convergence} is defined with respect to the mean dark matter density of the universe such that some regions are overdense (red), while others are underdense (blue). The critical curves are plotted in black for each realization. Density fluctuations associated with the wave interference of the dark matter profile in the main deflector halo cause small-scale features in the critical curve for $m_\psi \leq 10^{-20.5} \rm{eV}$. 
    We only generate density fluctuations in the vicinity of the critical curve, as this is the area where lensed images appear, but in principle they should exist throughout the entire halo. Note that the elliptical area in which fluctuations are rendered decreases with increasing $m_\psi$ to have a tractable number of fluctuations. The size (amplitude) of these density fluctuations varies proportionally (as the square root of) to the de Broglie wavelength associated with the particle mass. All realizations have $\Sigma_{\rm{sub}} = 0.025 \rm{kpc^{-2}}$, $\delta_{\rm{LOS}} = 1.0$, $\alpha = -1.9$, $\log_{10} (A_{\rm{fluc}}) = -1.3$ and $\gamma_\psi=1/3$.}
    \label{fig:proj_mass} 
\end{figure*}
\begin{figure*}
    \centering
    \includegraphics[width=0.95\linewidth]{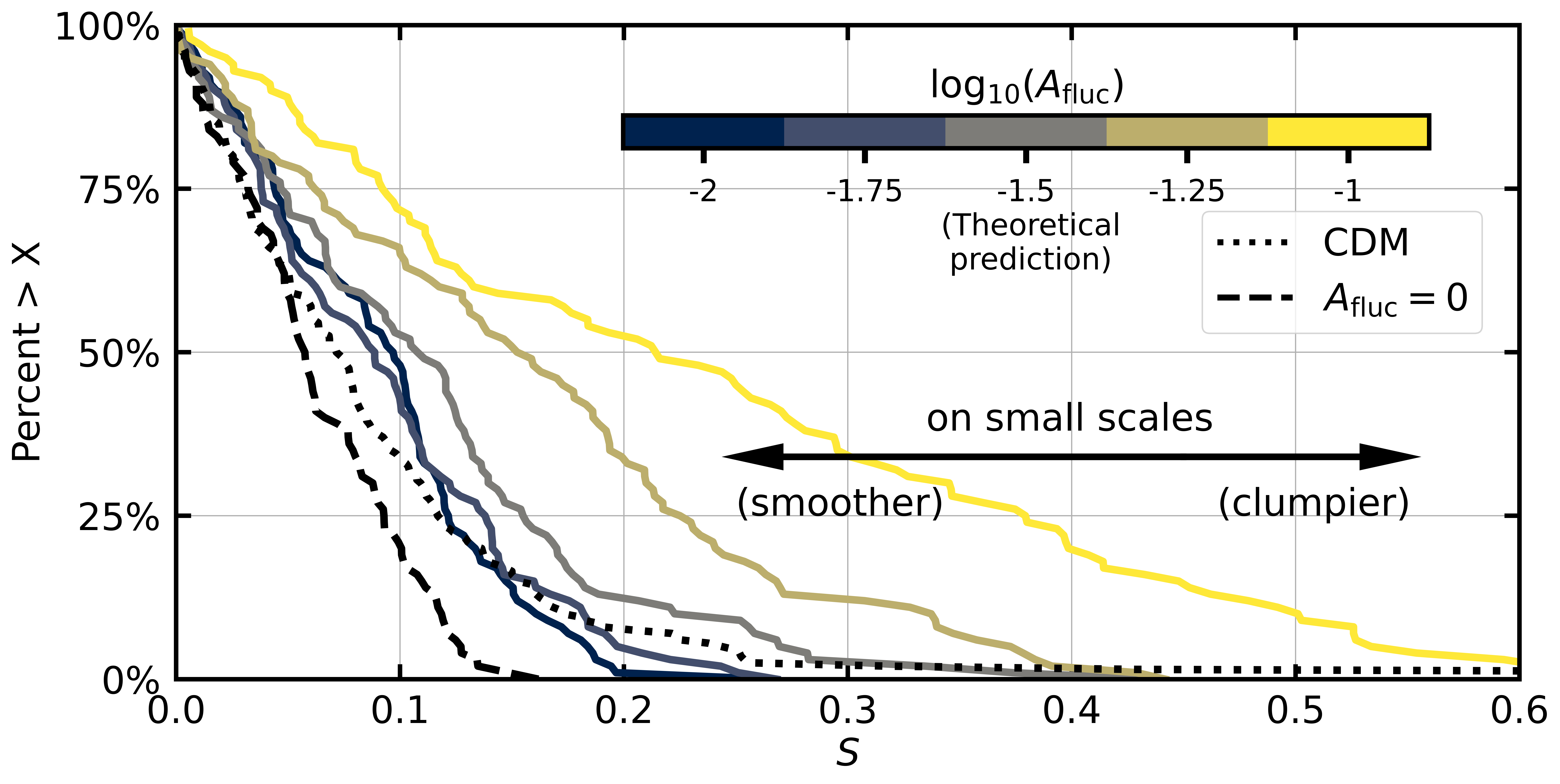}
    \caption{Cumulative distributions of the summary statistic $S$, Eq. \eqref{eqn:summarystat}, for various ULDM models with varying $A_{\rm{fluc}}$ at fixed particle mass $m_\psi=10^{-22}$ eV. The dashed black curve represents ULDM realizations without fluctuations ($A_{\rm{fluc}}=0$), whereas the solid colored curves represent ULDM realizations with fluctuations, for varying $A_{\rm{fluc}}$. The dotted black curve represents the distribution of $S$, assuming CDM. Small-scale density from larger values of $A_{\rm{fluc}}$ cause more frequent flux ratio anomalies, which results in longer tails in the cumulative distributions of the statistic. Values of $A_{\rm{fluc}}$ for which the the summary statistic distributions resemble CDM will have approximately equal likelihood, with respect to strong lensing data, even with very few dark matter halos present in the lens system due to the suppression of the halo mass function predicted by ULDM. We show how a wide range of $A_{\rm{fluc}}$ affect the data, but the value closest to the theoretical prediction, including baryons, ($\log_{10}(A_{\rm{fluc}})=-1.6\pm0.2$, see Section \ref{ssec:fluctuations}) is -1.5 (grey), and we include a prior that enforces this prediction in our main results.}
    \label{fig:sum_stat}
\end{figure*}

To explore how the constraints and the signal we extract from the data depends on the amplitude of the fluctuations, we implement a log-uniform prior on $A_{\rm{fluc}}$ around the best-fit value of $0.05$ used to create Figures \ref{fig:uldmcode} and \ref{fig:uldmcodepdfs}. In addition, we marginalize over a prior on $s$, which sets the size of a fluctuation relative to $\lambda_{\rm{dB}}$ and the number of fluctuations per units area. The prior on $s$ between 0.025 and 0.075 accounts for various factors that can change the size of a fluctuation for a given value of $m_{\psi}$, for example, a different central velocity dispersion of the host halo and central galaxy. 

The number of fluctuations generated in projection scales as $\frac{1}{\lambda_{\rm{dB}}^2}$, so the total number of fluctuations changes by a factor of $\sim 10^{12}$ across the log-uniform prior on $m_{\psi}$ between $[10^{-22.5},10^{-16.5}]$ eV. Direct implementation of this many individual lens profiles around each of the four images in ray-tracing computations is computationally intractable. To deal with this issue, we limit the total number of individual fluctuations rendered around each image to a number $n_{\rm{cut}}$. To approximately preserve the lensing properties of the density field when $n>n_{\rm{cut}}$, we can use the scaling of the variance $\delta \Sigma \propto \frac{1}{\sqrt{n}}$ and re-scale the amplitudes of the fluctuations by a factor of $\sqrt{\frac{n_{\rm{cut}}}{n}}$, where $n$ is the total number of fluctuations. We performed our analysis, discussed in Section \ref{sec:inferencemethod}, for different values of $n_{\rm{cut}}$ to determine the minimum value of $n_{\rm{cut}}$ at which the inference on $m_{\psi}$ converges, finding that $n_{\rm{cut}}=50,000$ is sufficient.

\subsection{Priors on model parameters}
\label{ssec:priors}
In this section, we provide a concise summary of definitions for the parameters describing the halo mass function, concentration-mass relation, halo density profiles, and the fluctuation density field. We also discuss physical assumptions attached to each prior. These parameters and their priors are also listed in Tables \ref{tab:params} and \ref{tab:nuisance_params}. 

We begin with the hyper-parameters listed in Table \ref{tab:params}. The distinction between these parameters and the nuisance parameters summarized in Table \ref{tab:nuisance_params} is that we do not combine information from multiple lenses to constrain the nuisance parameters, and instead marginalize over them on a lens-by-lens basis before multiplying likelihoods. On the other hand, we multiply likelihoods from different lenses to constrain the hyper-parameters listed below. 
\begin{itemize}
    \item $m_{\psi}$: We set a log-uniform prior on the particle mass $\log_{10}(m_\psi/\rm{eV})\in\mathcal{U}(-22.5,-16.5)$. Particle masses lighter than $10^{-22.5}\,\rm{eV}$ face stringent constraints the Cosmic Microwave Background \citep{Hlozek2015,Hlozek2017} using \textit{Planck} data \citep{Planck2015,Planck2018}, and constraints using the \textit{Hubble Ultra Deep Field} UV-luminosity function \citep{Bouwens_2015,Bozek_2015}  and the optical depth to reionization \citep{Spergel_2015}. The upper bound on $m_{\psi}$ is determined primarily by the halo mass function and concentration-mass relation. For particle masses $m_{\psi} > 10^{-17.5}\,\rm{eV}$, the abundance and density profiles of halos become practically indistinguishable from CDM, in the halo mass range that strong lensing is sensitive to (see Figures \ref{fig:SHMF} and \ref{fig:proj_mass}). Thus, extending the prior to $m_{\psi} = 10^{-16.5} \rm{eV}$ ensures that a region of parameter space we sample can be associated with CDM. 
    \item $\gamma_\psi\in\mathcal{U}(0.2,0.5)$: The prior on $\gamma_{\psi}$, which determines the size of the soliton core in ULDM halos, encompasses the range of theoretical uncertainty on this parameter based on the different values proposed to date. In addition, the specifics of the particle physics model and tidal disruption by the baryonic potential of the main deflector can alter the mass and size of soliton core \citep{Du2018,Glennon++22}. We account for these effects by marginalizing over $\gamma_{\psi}$ when quoting constraints on the particle mass. 
    \item $A_{\rm{fluc}}$: We assign a log-uniform prior on $A_{\rm{fluc}}$ between $10^{-3.5}$ and $10^{-0.5}$. While we eventually enforce a prior on this parameter based on our numerical simulations of ULDM halos (see Figures \ref{fig:uldmcode} and \ref{fig:uldmcodepdfs}), leaving $A_{\rm{fluc}}$ as a free parameter allows us to explore the interplay between halos and the fluctuations of the host halo profile in the lensing signal we measure. In Section \ref{ssec:posteriors}, we present results that assume a Gaussian prior on $\log_{10} A_{\rm{fluc}}$ with mean $-1.6$ and a variance of 0.2 dex based on calibrating our model against numerical solutions for host halo mass profiles in ULDM (see Section \ref{ssec:fluctuations}).  
    \item $\Sigma_{\rm{sub}}$: The normalization of the subhalo mass function encompasses a broad range of theoretical uncertainty associated with the efficiency of tidal stripping by baryons in the host halo potential. We account for the effects of tidal stripping by marginalizing over a uniform prior on $\Sigma_{\rm{sub}}$ between $0$ and $0.1 \rm{kpc^{-2}}$. More generally, we can associate $\Sigma_{\rm{sub}}$ with the overall contribution of subhalos to the lensing signal. Thus, the broad prior we assign to this parameter encompasses uncertainties associated with both the overall abundance and density profile of halos. 
    \item $\alpha$: The prior on $\alpha$, the logarithmic slope of the subhalo mass function, is motivated by N-body simulations of subhalos  \citep[e.g.][]{Springel++08,Fiacconi++16}.
    \item $\delta_{\rm{LOS}}$: The prior on $\delta_{\rm{LOS}}$, the amplitude of the field halo mass function relative to the prediction of the Sheth-Tormen halo mass function model, account for discrepancies between predictions of different theoretical models of the halo mass function below $10^{10}\,M_\odot$ \citep[e.g.][]{Despali++16}, cosmological model uncertainties in parameters such as $\sigma_8$ and $\Omega_M$, and the impact of baryonic matter on small-scale clustering \citep{Benson++20}.
\end{itemize}

We marginalize over nuisance parameters, listed in Table \ref{tab:nuisance_params}, if the parameters convey information that is specific to a particular lens. For example, as we expect each lensed quasar has a different source size (within the width of the prior) we marginalize over the source size before multiplying the likelihoods to constrain the particle mass.  

\begin{itemize}
    \item $M_{\rm{host}}$: We marginalize over a log-uniform prior on the host halo mass determined on a lens-by-lens basis based on the Einstein radius, lens and source redshift. The prior is based on correlations between these quantities and the stellar mass presented by \citet{Auger++10}, and between the stellar mass and host halo mass presented by \citet{Lagattuta++10}. We defer to \citet{Gilman2019a} for additional details.
    \item $s$: The parameter $s$ sets the size of an individual fluctuation in the projected mass profile of the host halo caused by the wave-like properties of the dark matter. The uncertainty in the size of a fluctuation reflects scatter in the velocity dispersion for each deflector that determines the characteristic length scale associated with the particle mass Equation \ref{eq:dB}.
    \item $a_4$: The parameter $a_4$ sets the amplitude of an octopole mass moment superimposed on the main deflector mass profile. When this additional component shares a common centroid and position angle with the elliptical power-law profile used to model the main deflector, it produces boxy or disky isophotes. The prior on $a_4$ is determined by observations of surface brightness contours of massive elliptical galaxies \citep{Bender1989}. 
    \item $\sigma_{\rm{src}}$: The source size sets the minimum angular scale where a deflection angle can affect an image magnification, and therefore determines the minimum halo mass detectable with the data. The source sizes of the nuclear narrow-line emission, radio, and CO 11-10 emission have sizes\footnote{We define the size as the full-width at half maximum, assuming a Gaussian profile.} between $1 - 60\,\rm{pc}$ \citep{MullerSanchez++11,Stacey18,chiba2005}. 
    \item $\gamma_{\rm{macro}}$: $\gamma_{\rm{macro}}$ sets the logarithmic profile slope of the main deflector mass profile. The prior between $1.95 - 2.2$ encompasses typical values for early-type galaxies \citep[e.g.][]{Gavazzi++07,Auger++10,Gilman++17}. 
    \item $\gamma_{\rm{ext}}$: The prior on the magnitude of the external shear is determined on a lens-by-lens basis by running the inference pipeline with the shear left as a free parameter, and determining what ranges of $\gamma_{\rm{ext}}$ can reproduce the observed flux ratios. 
    \item $\delta_{x,y}$: We add astrometric uncertainties to the image positions. The amount of uncertainty is determined by the measurements of the image positions and flux ratios \citep{Nierenberg2019}.
    \item $\delta f$: We add measurement errors to the image fluxes, or in some cases, to the model flux ratios, depending on which quantity has published errorbars \citep{chiba2005,Nierenberg2019}.  
    
\end{itemize}

\section{Understanding the effect of structure formation with ULDM on image flux ratios}
\label{sec:summary_stat}

Before discussing the results of the full forward modeling and Bayesian inference on the particle mass $m_{\psi}$, we investigate the effects of ULDM halos and density fluctuations on flux ratios to build intuition that will aid the interpretation of our main results. To this end, Figure \ref{fig:proj_mass} shows eight projected mass distributions of dark matter structure generated using the parameterization for the mass function, concentration-mass relation, and fluctuations throughout the host halo density profile presented in the previous section. Each panel shows a map of the effective multi-plane convergence in substructure, $\kappa_{\rm{effective(halo)}}$, given by
\begin{ceqn}
\begin{equation}
\label{eqn:kappaeffective}
    \kappa_{\rm{effective(halo)}} \equiv \frac{1}{2} {\mathbf{\nabla} \cdot \boldsymbol{\alpha}} - \kappa_{\rm{macro}},
\end{equation}
\end{ceqn}
where $\boldsymbol{\alpha}$ is the multi-plane deflection field, and $\kappa_{\rm{macro}}$ is a smooth mass profile used to model the main deflector. The effective multi-plane convergence in substructure is a two-dimensional representation of a full three-dimensional population of halos, subhalos, and in the case of ULDM, quantum fluctuations of projected mass profile of the host halo. The definition of $\kappa_{\rm{effective}(halo)}$ includes non-linear effects associated with multi-plane lensing, such that halos along the line of sight appear distorted in the direction tangent to the critical curve. 

Each panel in Figure \ref{fig:proj_mass} depicts a realization of ULDM structure for a lens at redshift $z=0.5$ and a source at $z = 2.0$, with values for the hyper-parameters introduced in the previous paragraphs of $\Sigma_{\rm{sub}} = 0.025\,\rm{kpc^{-2}}$, $\delta_{\rm{LOS}} = 1.0$, $\alpha = -1.9$, $\log_{10} (A_{\rm{fluc}}) = -1.3$ and $\gamma_\psi=1/3$. We include, as a point of comparison, a CDM realization in the bottom right panel. Beginning in the top left with a $10^{-22}\,\rm{eV}$ particle, very few halos or subhalos exist in the lens system, but fluctuations in the background density of the host halo appear prominently and produce visible distortions in the critical curve, shown in black. Increasing the particle mass, the fluctuation amplitude decays as $m_{\psi}^{-1/2}$, such that the fluctuations become nearly imperceptible for $m_{\psi} > 10^{-20.5}\,\rm{eV}$, while progressively more halos appear in the lens system as $M_{1/2}$ decreases as $m_{\psi}^{-4/3}$. For particle masses greater than $m_\psi\sim10^{-17}\,\rm{eV}$, the halo populations are practically indistinguishable from CDM in the mass range $10^6-10^{10} M_{\odot}$ relevant for substructure lensing. 
\begin{figure}
    \centering
    \includegraphics[width=0.45\textwidth]{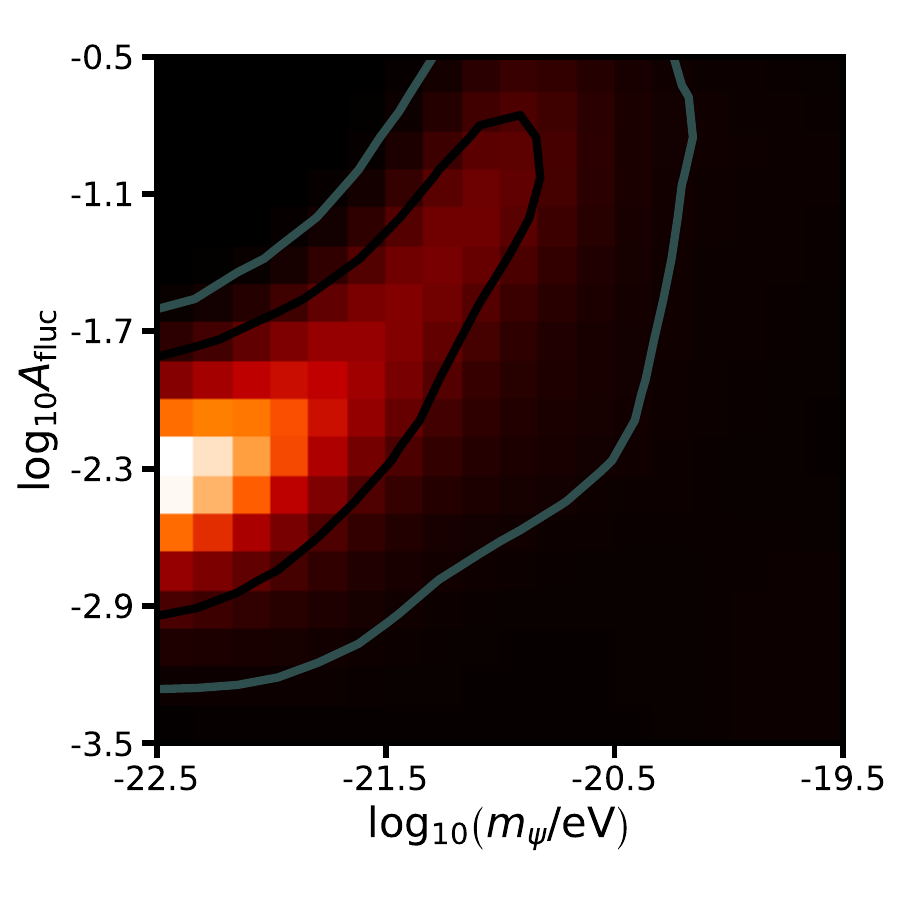}
    \caption{Constraints from eleven lenses on the particle mass $m_\psi$ and the fluctuation amplitude $A_{\rm{fluc}}$, with no halos included in the lens model. Fluctuation amplitudes $\log_{10}(A_{\rm{fluc}}) > -1.5$ and light particles $m_{\psi} < 10^{-21.5}\,\rm{eV}$ are ruled out because they impart too much perturbation to image flux ratios (see also Figure \ref{fig:sum_stat}). On the other hand, fluctuation amplitudes $\log_{10}(A_{\rm{fluc}}) < -3.0$ and more massive particles $m_{\psi} > 10^{-20}\,\rm{eV}$ are ruled out because the fluctuations are strongly suppressed in this regime.}
    \label{fig:flucsonly}
\end{figure}

The rich diversity of structure formation outcomes shown in Figure \ref{fig:proj_mass} implies that the presence of fluctuations from wave inference of the the dark matter can, at least to some extent, compensate for the relative lack of halos in ULDM models with $m_{\psi} \leq 10^{-20.5}$ eV. Certain values of $A_{\rm{fluc}}$ could cause enough perturbation to flux ratios to explain the data, despite the fact that very few halos exist in the lens system. 

We can explore the relative impact of fluctuations to halos, as determined by the fluctuation amplitude $A_{\rm{fluc}}$, by computing distributions of the summary statistic $S$ defined in Equation \ref{eqn:summarystat}. Figure \ref{fig:sum_stat} shows cumulative distributions of $S$ for different choices of $A_{\rm{fluc}}$, assuming $m_{\psi} = 10^{-22}\,\rm{eV}$. To compute the statistic, we compute reference flux ratios $f_{\rm{data(i)}}$ from a smooth lens model with no halos or fluctuations present in the lens system. Thus, cumulative distributions of $S$ with long tails indicate frequent and strong perturbations to image flux ratios, while a cumulative distribution that rapidly drops to zero along the $x$-axis corresponds to infrequent and/or small perturbations to the data. The non-zero values of $S$ for the dashed curve, which includes no fluctuations, represents the variation of image flux ratios that results from marginalizing over the mass profile of the main deflector (the logarithmic profile slope, external shear, boxyness and diskyness, etc.) as well as the relatively few remaining halos when $m_{\psi} = 10^{-22}\,\rm{eV}$. 

The colored curves in Figure \ref{fig:sum_stat} correspond to different values of $A_{\rm{fluc}}$, while the dotted line shows a distribution of $S$ for CDM. Adding fluctuations to the main deflector halo, the amount of perturbation predicted by ULDM theories can far exceed that predicted by CDM, depending on the value of $A_{\rm{fluc}}$. Figure \ref{fig:sum_stat} clearly demonstrates that both halos and fluctuations can perturb image flux ratios. In fact, both theoretical frameworks predict similar amounts of perturbation provided $A_{\rm{fluc}} \sim 10^{-1.75}-10^{-2}$. Thus, based on Figure \ref{fig:sum_stat}, we can expect fluctuation amplitudes $A_{\rm{fluc}} \sim 10^{-2}$ and $m_{\psi} \sim 10^{-22}\rm{eV}$ will not be ruled out by the data. To explore this possibility rigorously, and to disentangle the effects of density fluctuations of the host halo profile from the effects of halos in the lens model, we now apply the full forward modeling pipeline reviewed in Section \ref{sec:inferencemethod} to the structure formation model presented in Section \ref{sec:model}.

\section{Results}
\label{sec:results}
\begin{figure*}
    \centering
    \includegraphics[width=\linewidth]{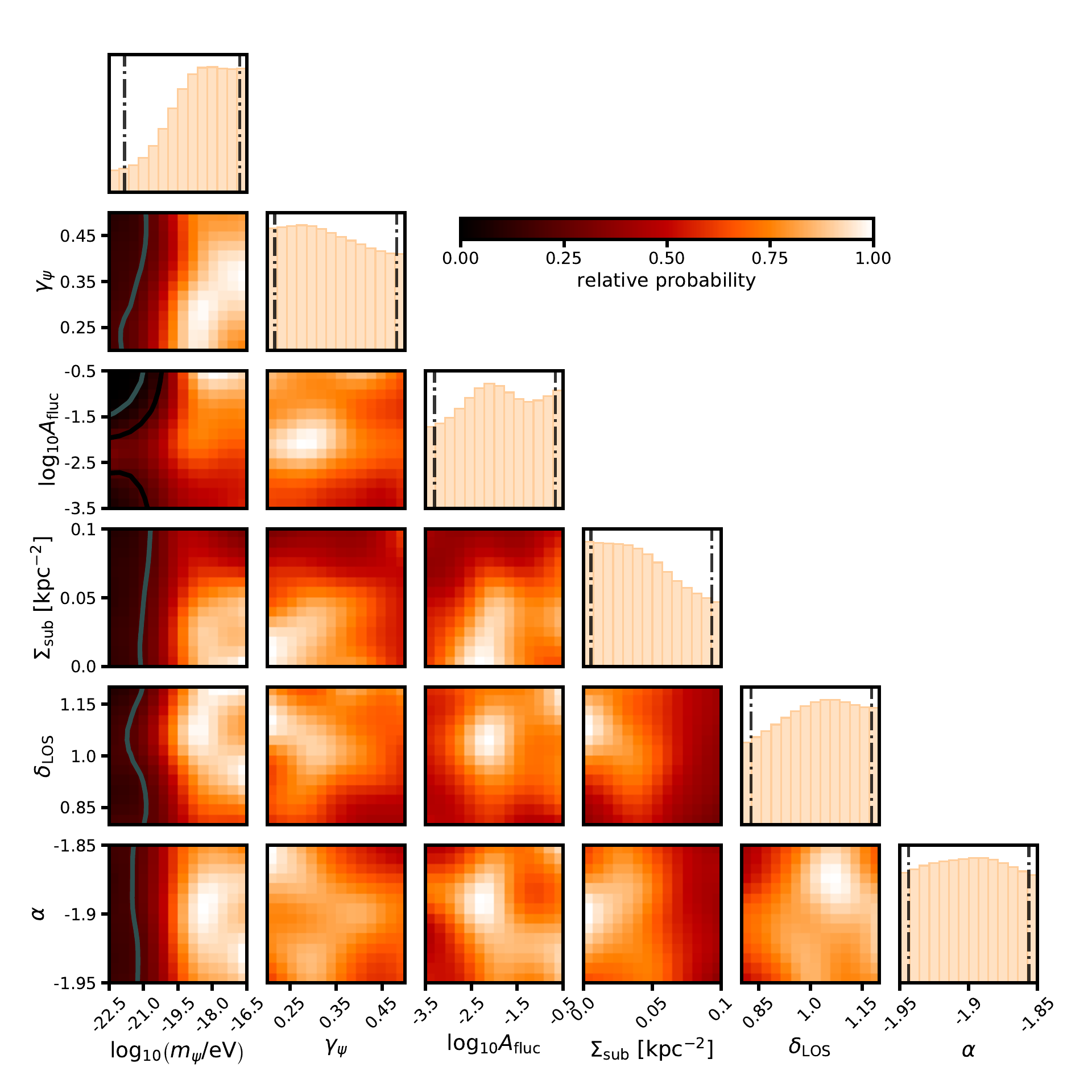}
    \caption{Joint posterior distribution for ultra-light dark matter parameters. We display the particle mass $m_\psi,$ the core radius-halo mass power lax exponent $\gamma_\psi,$ the fluctuation amplitude $A_{\rm{fluc}}$, and the normalization of the subhalo mass function $\Sigma_{\rm{sub}},$ the rescaling factor for the line of sight Sheth-Thormen mass function $\delta_{\rm{LOS}}$ and the logarithmic slope of the subhalo mass function $\alpha$. Vertical dotted-dashed lines on the marginal distributions denote 95$\%$ confidence intervals and black (grey) contours denote 68$\%$ (95$\%$) confidence intervals.}
    \label{fig:corner}
\end{figure*}

\begin{figure*}
    \centering
    \includegraphics[width=\linewidth]{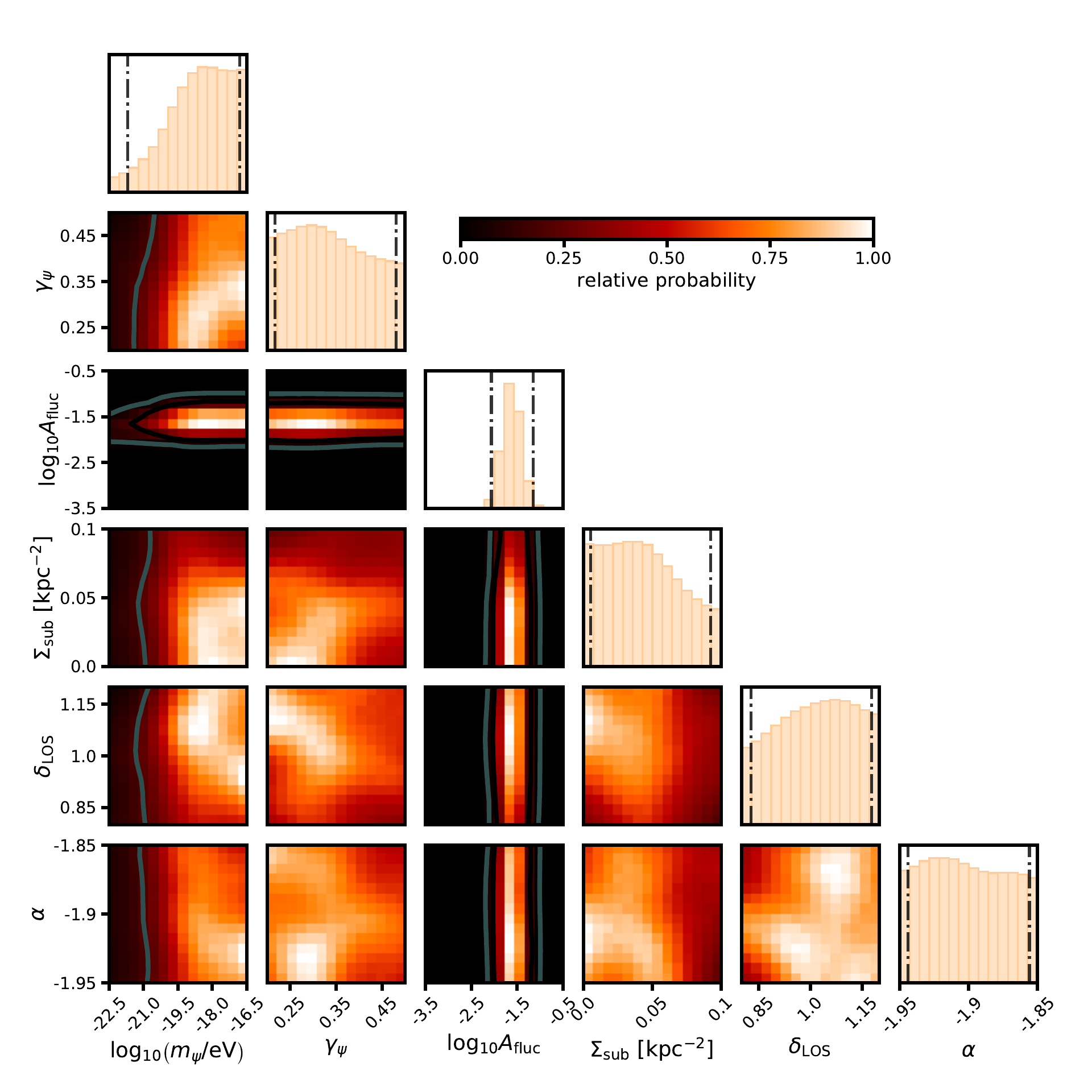}
    \caption{The same inference as shown in Figure \ref{fig:corner}, but including a prior on the fluctuation amplitude $A_{\rm{fluc}}$ derived from numerical simulations of structure formation in ULDM (see Section \ref{ssec:fluctuations}).}
    \label{fig:prior}
\end{figure*}

We combine the inference method and dataset described in Section \ref{sec:inferencemethod} with the ULDM structure formation model detailed in Section \ref{sec:model} to obtain constraints on the particle mass $m_\psi$. In Section \ref{ssec:flucsonly}, we begin by considering a simplified model with no halos and only quantum fluctuations included in the lens model to demonstrate how fluctuations in the host halo mass profile can affect strong lensing observables. We present joint constraints on the particle mass and fluctuation amplitude for this scenario. In Section \ref{ssec:posteriors}, we present the joint posterior distribution for ULDM parameters from the eleven quad lenses in our sample, including halos and fluctuations. Folding in theoretical predictions for the fluctuation amplitude $A_{\rm{fluc}}$, we derive constraints on the particle mass $m_\psi$, and compare the constraints to a model in which no fluctuations, and only halos, are included in the lens model. 

\subsection{Isolating the lensing signal from fluctuations}\label{ssec:flucsonly}

We may better understand the effect of the fluctuation amplitude on our particle mass constraints by isolating the the signal from density fluctuations. Figure \ref{fig:flucsonly} presents our joint constraint on $m_\psi$ and $A_{\rm{fluc}}$, assuming no halos exist in the lens model. As both the amplitude and size of individual fluctuations scale inversely with $m_{\psi}$, the effect of the fluctuations becomes increasingly suppressed as $m_{\psi}$ increases. Thus, with only fluctuations included in the model, the data rules out particle masses greater than approximately $10^{-20.5}\,\rm{eV}$ because these models predict too little flux ratio perturbation in the data. 

For lighter particle masses, $m_{\psi} < 10^{-20.5}\,\rm{eV}$, the fluctuations have large sizes and amplitudes, and thus their presence has a significant impact on the data. In particular, increasing $A_{\rm{fluc}}$ increases the central density of individual fluctuations, boosting their lensing efficiency. This results in too much perturbation, and thus the data rules out parameter space that occupies the upper-left section of Figure \ref{fig:flucsonly}. Similarly, decreasing $A_{\rm{fluc}}$ again results in too little perturbation, and thus the data disfavors regions of parameter space that occupy the bottom left of Figure \ref{fig:flucsonly}. 

The likelihood contours in Figure \ref{fig:flucsonly} track the region of parameter space where the fluctuations-only model can explain the data. Curiously, these likelihood contours approximately follow $A_{\rm{fluc}}\sim m_\psi^{1/2}$. By Equation \ref{eqn:flucamp}, this corresponds to a fluctuation amplitude independent of the particle mass, since the characteristic amplitude of the density fluctuations, $\sqrt{\langle \delta \kappa^2 \rangle}$ (see Equation \ref{eqn:flucamp}), itself scales as $m_\psi^{-1/2}$. For a $10^{-22}\,\rm{eV}$ particle, this leads to an amplitude of $\sqrt{\langle \delta \kappa^2 \rangle} = 0.0029_{-0.0017}^{+0.0037}$. \footnote{Although it is not within the scope this work, we note that one could interpret the likelihood in Figure \ref{fig:flucsonly} in terms of the power spectrum of dark substructure in strong lenses \citep[e.g.][]{Hezaveh++16,DiazRivero++18,Cyr-Racine++19}, as $A_{\rm{fluc}}$ and $m_{\psi}$ jointly determine the amplitude and size of projected density fluctuations in the lens.}

Analyzing the data with only fluctuations included in the model provides a useful illustration of how the fluctuations impact the data. However, the model in terms of only $m_{\psi}$ and $A_{\rm{fluc}}$ does not adhere to the predictions of any physically-motivated dark matter theory proposed to date. The next section presents constraints from our complete ULDM model, which includes halos and subhalos, in addition to the fluctuations.

\subsection{Constraints on the particle mass}
\label{ssec:posteriors}

The joint posterior distribution we infer from our analysis for $m_\psi,\,\gamma_\psi,A_{\rm{fluc}},\,\Sigma_{\rm{sub}},\,\delta_{\rm{LOS}}$ and $\alpha$ is shown in Figure \ref{fig:corner}. Without imposing a theoretically-motivated prior on $A_{\rm{fluc}}$, we see clear covariance between $m_\psi$ and $A_{\rm{fluc}}$ in the full model (halos and fluctuations). The posterior for $m_\psi$ demonstrates that we disfavor light particle masses $(m_\psi<10^{-21}$ eV) provided that $\log_{10}(A_{\rm{fluc}})\lesssim-3$ or  $\log_{10}(A_{\rm{fluc}})\gtrsim-1.5$. However, as one can infer from comparing the curves in Figure \ref{fig:sum_stat}, the data does not strongly disfavor $m_{\psi} \sim 10^{-22}\,\rm{eV}$ provided $A_{\rm{fluc}} \sim 10^{-2}$. 

\begin{figure}
    \includegraphics[trim=3cm 0.75cm 3cm
	0.5cm,width=0.45\textwidth]{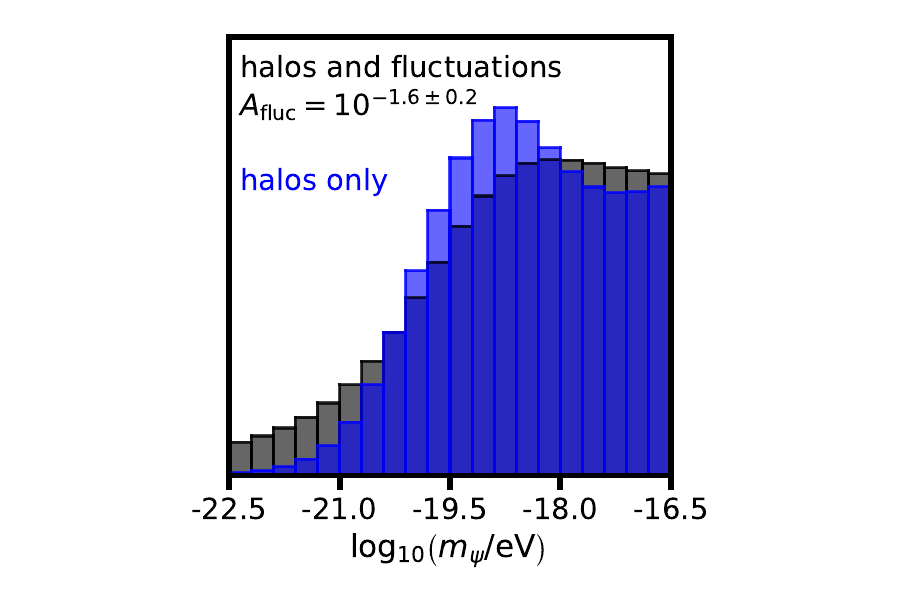}
    \caption{Marginal distribution of the particle mass $m_{\psi}$ for the full structure formation model that includes halos and fluctuations shown in Figure \ref{fig:prior} (black), and a model without fluctuations and only halos included in the lens model (blue). Including fluctuations in the projected mass profile of the host halo increases the relative likelihood of CDM to $\log_{10}\left(m_{\psi} \ \rm{eV}\right) \in \left[-22.5, -22.25\right]$ by a factor of 12.}
    \label{fig:noflucspdf}
\end{figure}

We now fold in a more informative prior on the fluctuation amplitude based on our numerical simulations of the host halo density profile with the software developed by \citet{Yavetz++22}. As discussed in Section \ref{ssec:fluctuations}, we can match our model to the simulations assuming $\log_{10}(A_{\rm{fluc}})=-1.6\pm0.2,$ accounting for the presence of baryonic mass projected near the Einstein radius. Figure \ref{fig:prior} shows a similar joint posterior distribution to Figure \ref{fig:corner}, while imposing a Gaussian prior $\mathcal{N}(-1.6,0.2)$ on $\log_{10}(A_{\rm{fluc}})$. Accounting for theoretical predictions, we obtain a lower bound for the ultra-light boson mass: $m_\psi>3.2\times10^{-22}$ eV at $95\%$ confidence, although this constraint depends fairly strongly on the width of the prior between $10^{-22.5} - 10^{-16.5}\,\rm{eV}$. As a more robust metric we can quote a relative likelihood, defined as the probability that $m_{\psi}$ has a value in a logarithmically-spaced interval relative CDM. We define the likelihood of CDM as the probability that $\log_{10}(m_\psi/\rm{eV})\in[-16.75, -16.5]$, because strong lensing is practically insensitive to differences between ULDM and CDM for particle masses in this range. Our constraints expressed in terms of relative likelihoods favor CDM over ULDM with a particle mass in the range $\log_{10}(m_\psi/\rm{eV})\in[-22.5,-22.25], [-22.25,-22.0],[-22.0,-21.75],[-21.75,-21.5]$ with odds of 8:1, 7:1, 6:1, and 4:1. 

While the data clearly penalizes ULDM models with particle masses $m_{\psi} < 10^{-21.5}\,\rm{eV},$ the constraints are significantly weaker than if we had based our analysis only on the impact of ULDM on the halo mass function, the concentration-mass relation, and halo density profiles. To illustrate this, we run our inference pipeline on a model identical to the one presented in Section \ref{sec:model}, but with $A_{\rm{fluc}} = 0$, removing the fluctuations from the model. Figure \ref{fig:noflucspdf} compares the marginal likelihood of $m_{\psi}$ shown in Figure \ref{fig:prior}, which includes our theoretically-motivated prior on the fluctuation amplitude, with the inference on $m_{\psi}$ that results from omitting the fluctuations. Accounting for only halos in the analysis (blue distribution in Figure \ref{fig:noflucspdf}), the likelihood of CDM relative to  $\log_{10}(m_\psi/\rm{eV})\in[-22.5,-22.25], [-22.25,-22.0],[-22.0,-21.75],[-21.75,-21.5]$ is 98:1, 48:1, 26:1 and 18:1, respectively. Thus, omitting the quantum fluctuations of the host dark matter halo profile from the lensing analyses causes one to conflate their impact on the data with perturbations by dark matter halos, biasing inferences on the particle mass. 

\section{Discussion and conclusions}
\label{sec:discussion}

We present an analysis of ultra-light dark matter (ULDM) and constraints on the ULDM particle mass using a sample of eleven quadruply-imaged quasars, using the Bayesian inference methodology developed by \citep{Gilman2019c,Gilman2019a}. The structure formation model we implement to study ULDM accounts for suppressed halo abundance and concentration in ULDM frameworks, the presence of a central soliton core in individual halos, and the quantum fluctuations of the host halo density profile arising from wave interference effects. We summarize our main results as follows:
\begin{itemize}
    \item When taking theoretical predictions for the fluctuation amplitude into account, the data favors CDM over ULDM with relative likelihoods of $\log_{10}(m_\psi/\rm{eV})\in[-22.5,-22.25]$ $[-22.25,-22.0],[-22.0,-21.75], [-21.75,-21.5]$ with a likelihood ratio of 8:1, 7:1, 6:1, and 4:1. We express these results in terms of relative likelihoods, instead of confidence intervals, because these metrics do not depend on the prior assigned to the particle mass in our analysis. 
    \item Constraints on ULDM from strong gravitational lensing depend on the quantum fluctuations that exist in ULDM halo profiles. These fluctuations, associated with wave interference effects in ULDM, cause significant perturbation to flux ratios. Specifically, constraints on the ultra-light boson mass $m_\psi$ depend on the fluctuation amplitude $A_{\rm{fluc}}$ of the host dark matter halo, which should be accounted for in a strong lensing analysis of ULDM.
\end{itemize}

This paper shows that strong gravitational lensing by galaxies provides a powerful astrophysical test of ULDM. In particular, we have shown that lensing provides a direct gravitational probe of small-scale structure generated through wave inference effects in ULDM. For lensing-based inferences on the amount of small-scale structure in galactic halos, the density fluctuations associated with the wave interference compensate for the suppression of small-scale structure in ULDM that results from the cutoff in the linear matter power spectrum. This ultimately leads to weaker constraints on the ULDM particle mass than one would obtain by considering only the effects of halos on the data. A different way of phrasing this result is that flux ratios do not distinguish between a small-scale density fluctuation associated with a halo, and a wave-like density fluctuation of the host halo density profile. Adding additional information that distinguishes between different angular scales, such as gravitational imaging data \citep[e.g.][]{Vegetti2014,Hezaveh++16} or flux ratios measured with more compact source sizes \citep{Gilman2021}, could help distinguish fluctuations (where the relevant angular scale is set by the de Broglie wavelength relative to the source size) from halos (where the relevant angular scale is set by the halo mass relative to the source size). 

This paper arrives on the heels of several studies that place stringent limits on the particle mass from the Lyman-$\alpha$ forest \citep{Keir2021}: $m_\psi>2\times10^{-20}$ eV at 95\% confidence, stellar orbits in ultra-faint dwarf galaxies \citep{Dalal++22}: $m_\psi>3\times10^{-19}$ eV at 99\% confidence, black hole superradiance \citep{Davoudiasl_Denton19}: ruling out $m_\psi\in(2.9,4.6)\times10^{-21}$ eV and Jeans analysis for dwarf spheroidal galaxies \citep{Chen++2017}: favoring $m_\psi=1.18^{+0.28}_{-0.24}\times10^{-22}$ (using \citet{Walker++07}) and $m_\psi=1.18^{+0.35}_{-0.33}\times10^{-22}$ (using \citet{Walker++09}) at 95\% confidence. Additionally, \citet{DellaMonica++22} recently demonstrated that future observations of the orbital motion of the S2 star will place an upper limit on the particle mass. The key distinguishing feature of the results we present, relative to these other works, is that lensing provides a direct gravitational probe of the dark matter structure predicted by ULDM, rather than using a proxy observable quantity, such as the Lyman-$\alpha$ flux power spectrum, stellar orbits or dSph galaxies, to infer properties of the unobservable dark matter. This means that lensing depends on different modeling assumptions and sources of systematic error.

An increase in the observed strong lensing dataset by an order of magnitude \citep{oguri2010,treu2018} will lead to more stringent bounds on the particle mass. In addition, new measurements of image flux ratios from the James Webb Space Telescope through JWST GO-2046 \citep{Nierenberg++21} in mid-infrared wavelengths will provide a more sensitive probe of dark substructure in a sample of approximately thirty strong lenses. Relative to the narrow-line flux ratios that constitute the majority of the data analyzed in this work, the mid-IR emission comes from a more compact area around the background quasar. This increases the sensitivity of the data to smaller deflection angles, effectively pushing the resolution of the data to lower halo masses. 

The constraints in this paper can also be considered in the context of ultralight vector dark matter (VDM). \citet{Amin++22} studied the small scale structure differences between ULDM and VDM. They determined that the amplitude of the fluctuations in VDM, relative to ULDM, is reduced by a factor of 1/3 due to a decrease in interference between ultralight vector bosons. Generally, fluctuation amplitudes for a spin $s$-bosonic field would decrease by a factor of $1/(2s+1).$ Since VDM reduces perturbations to flux ratios from fluctuations, the lower limits presented in this paper are also conservative bounds for VDM.

Both warm dark matter and ULDM exhibit a truncated matter power spectrum, and a characteristic half-mode mass below which the abundance and density profiles become suppressed, relative to CDM. \citet{Schutz2020} used this fact to translate bounds on warm dark matter models inferred from strong lenses \citep{Gilman2019a} and stellar streams \citep{Banik++21} to constraints on the ULDM particle mass, inferring $m_\psi>2.1\times10^{-21}$ eV at $95\%$ confidence. The results of this paper demonstrate that strong lensing constraints on ULDM depend on the properties of the fluctuations of the host halo, and thus mapping a constraint from warm dark matter obtained through lensing to ultra-light dark matter gives a misleading result. This conclusion is similar to the one reached by \citet{Dalal++21}, who show that the wave-like phenomena that distinguishes ULDM from WDM also affects the small-scale structure of stellar streams. 

\section*{Acknowledgements}

We thank the anonymous referee for their careful reading of our manuscript and their insightful comments. We also thank Andrew Benson, Simon Birrer, and Lam Hui for feedback on this work. Finally, we thank Anowar Shajib for sharing results from \citet{Shajib++21}.

AL acknowledges support from a Natural Sciences and Engineering Research Council of Canada (NSERC) Undergraduate Summer Research Award. DG was partially supported by a HQP grant from the McDonald Institute (reference number HQP 2019-4-2). XL is supported by NSERC, funding
reference \#CITA 490888-16 and the Jeffrey L. Bishop
Fellowship. Research at Perimeter Institute is supported in
part by the Government of Canada through the Department of
Innovation, Science and Economic Development Canada and the
Province of Ontario through the Ministry of Colleges and
Universities. AL and JB acknowledge financial support from an Ontario Early Researcher Award (ER16-12-061). DG and JB acknowledge financial support from NSERC (funding reference number RGPIN-2020-04712). XD acknowledges support from NASA ATP grant 17-ATP17-0120.

This work consumed approximately 300,000 CPU hours
distributed across two computing clusters. First, we performed computations on the Niagara supercomputer at the SciNet HPC Consortium \citep{Loken2010,Ponce2019}. SciNet is funded by: the Canada Foundation for Innovation; the Government of Ontario; Ontario Research Fund -
Research Excellence; and the University of Toronto. Second,
we used computational and storage services associated with
the Hoffman2 Shared Cluster provided by the UCLA Institute for Digital Research and Education’s Research Technology Group.

%%%%%%%%%%%%%%%%%%%%%%%%%%%%%%%%%%%%%%%%%%%%%%%%%%
\section*{Data Availability}

Data associated with this paper are available upon reasonable request from the corresponding author.



\bibliographystyle{mnras}
\bibliography{paper} % if your bibtex file is called example.bib

\begin{thebibliography}{}
\makeatletter
\relax
\def\mn@urlcharsother{\let\do\@makeother \do\$\do\&\do\#\do\^\do\_\do\%\do\~}
\def\mn@doi{\begingroup\mn@urlcharsother \@ifnextchar [ {\mn@doi@}
  {\mn@doi@[]}}
\def\mn@doi@[#1]#2{\def\@tempa{#1}\ifx\@tempa\@empty \href
  {http://dx.doi.org/#2} {doi:#2}\else \href {http://dx.doi.org/#2} {#1}\fi
  \endgroup}
\def\mn@eprint#1#2{\mn@eprint@#1:#2::\@nil}
\def\mn@eprint@arXiv#1{\href {http://arxiv.org/abs/#1} {{\tt arXiv:#1}}}
\def\mn@eprint@dblp#1{\href {http://dblp.uni-trier.de/rec/bibtex/#1.xml}
  {dblp:#1}}
\def\mn@eprint@#1:#2:#3:#4\@nil{\def\@tempa {#1}\def\@tempb {#2}\def\@tempc
  {#3}\ifx \@tempc \@empty \let \@tempc \@tempb \let \@tempb \@tempa \fi \ifx
  \@tempb \@empty \def\@tempb {arXiv}\fi \@ifundefined
  {mn@eprint@\@tempb}{\@tempb:\@tempc}{\expandafter \expandafter \csname
  mn@eprint@\@tempb\endcsname \expandafter{\@tempc}}}

\bibitem[\protect\citeauthoryear{{Abazajian} \& {Kusenko}}{{Abazajian} \&
  {Kusenko}}{2019}]{Abazajian++19}
{Abazajian} K.~N.,  {Kusenko} A.,  2019, \mn@doi [\prd]
  {10.1103/PhysRevD.100.103513}, \href
  {https://ui.adsabs.harvard.edu/abs/2019PhRvD.100j3513A} {100, 103513}

\bibitem[\protect\citeauthoryear{{Abbott} \& {Sikivie}}{{Abbott} \&
  {Sikivie}}{1983}]{Abbott1982}
{Abbott} L.~F.,  {Sikivie} P.,  1983, \mn@doi [Physics Letters B]
  {10.1016/0370-2693(83)90638-X}, \href
  {https://ui.adsabs.harvard.edu/abs/1983PhLB..120..133A} {120, 133}

\bibitem[\protect\citeauthoryear{{Amendola} \& {Barbieri}}{{Amendola} \&
  {Barbieri}}{2006}]{Amendola06}
{Amendola} L.,  {Barbieri} R.,  2006, \mn@doi [Physics Letters B]
  {10.1016/j.physletb.2006.08.069}, \href
  {https://ui.adsabs.harvard.edu/abs/2006PhLB..642..192A} {642, 192}

\bibitem[\protect\citeauthoryear{{Amin}, {Jain}, {Karur}  \& {Mocz}}{{Amin}
  et~al.}{2022}]{Amin++22}
{Amin} M.~A.,  {Jain} M.,  {Karur} R.,   {Mocz} P.,  2022, arXiv e-prints,
  \href {https://ui.adsabs.harvard.edu/abs/2022arXiv220311935A} {p.
  arXiv:2203.11935}

\bibitem[\protect\citeauthoryear{{Amorisco} et~al.,}{{Amorisco}
  et~al.}{2022}]{Amorisco++22}
{Amorisco} N.~C.,  et~al., 2022, \mn@doi [\mnras] {10.1093/mnras/stab3527},
  \href {https://ui.adsabs.harvard.edu/abs/2022MNRAS.510.2464A} {510, 2464}

\bibitem[\protect\citeauthoryear{{Arvanitaki}, {Dimopoulos}, {Dubovsky},
  {Kaloper}  \& {March-Russell}}{{Arvanitaki} et~al.}{2010}]{arvanitaki2010}
{Arvanitaki} A.,  {Dimopoulos} S.,  {Dubovsky} S.,  {Kaloper} N.,
  {March-Russell} J.,  2010, \mn@doi [\prd] {10.1103/PhysRevD.81.123530}, \href
  {https://ui.adsabs.harvard.edu/abs/2010PhRvD..81l3530A} {81, 123530}

\bibitem[\protect\citeauthoryear{{Auger}, {Treu}, {Bolton}, {Gavazzi},
  {Koopmans}, {Marshall}, {Moustakas}  \& {Burles}}{{Auger}
  et~al.}{2010}]{Auger++10}
{Auger} M.~W.,  {Treu} T.,  {Bolton} A.~S.,  {Gavazzi} R.,  {Koopmans}
  L.~V.~E.,  {Marshall} P.~J.,  {Moustakas} L.~A.,   {Burles} S.,  2010,
  \mn@doi [\apj] {10.1088/0004-637X/724/1/511}, \href
  {https://ui.adsabs.harvard.edu/abs/2010ApJ...724..511A} {724, 511}

\bibitem[\protect\citeauthoryear{{Banik}, {van den Bosch}, {Tremmel}, {More},
  {Despali}, {More}, {Vegetti}  \& {McKean}}{{Banik} et~al.}{2019}]{Banik++19}
{Banik} U.,  {van den Bosch} F.~C.,  {Tremmel} M.,  {More} A.,  {Despali} G.,
  {More} S.,  {Vegetti} S.,   {McKean} J.~P.,  2019, \mn@doi [\mnras]
  {10.1093/mnras/sty3267}, \href
  {https://ui.adsabs.harvard.edu/abs/2019MNRAS.483.1558B} {483, 1558}

\bibitem[\protect\citeauthoryear{{Banik}, {Bovy}, {Bertone}, {Erkal}  \& {de
  Boer}}{{Banik} et~al.}{2021a}]{Banik++21a}
{Banik} N.,  {Bovy} J.,  {Bertone} G.,  {Erkal} D.,   {de Boer} T.~J.~L.,
  2021a, \mn@doi [\mnras] {10.1093/mnras/stab210}, \href
  {https://ui.adsabs.harvard.edu/abs/2021MNRAS.502.2364B} {502, 2364}

\bibitem[\protect\citeauthoryear{{Banik}, {Bovy}, {Bertone}, {Erkal}  \& {de
  Boer}}{{Banik} et~al.}{2021b}]{Banik++21}
{Banik} N.,  {Bovy} J.,  {Bertone} G.,  {Erkal} D.,   {de Boer} T.~J.~L.,
  2021b, \mn@doi [\jcap] {10.1088/1475-7516/2021/10/043}, \href
  {https://ui.adsabs.harvard.edu/abs/2021JCAP...10..043B} {2021, 043}

\bibitem[\protect\citeauthoryear{{Bender}, {Surma}, {Doebereiner},
  {Moellenhoff}  \& {Madejsky}}{{Bender} et~al.}{1989}]{Bender1989}
{Bender} R.,  {Surma} P.,  {Doebereiner} S.,  {Moellenhoff} C.,   {Madejsky}
  R.,  1989, \aap, \href
  {https://ui.adsabs.harvard.edu/abs/1989A&A...217...35B} {217, 35}

\bibitem[\protect\citeauthoryear{{Benson}}{{Benson}}{2012}]{Benson12}
{Benson} A.~J.,  2012, \mn@doi [\na] {10.1016/j.newast.2011.07.004}, \href
  {https://ui.adsabs.harvard.edu/abs/2012NewA...17..175B} {17, 175}

\bibitem[\protect\citeauthoryear{{Benson}}{{Benson}}{2020}]{Benson++20}
{Benson} A.~J.,  2020, \mn@doi [\mnras] {10.1093/mnras/staa341}, \href
  {https://ui.adsabs.harvard.edu/abs/2020MNRAS.493.1268B} {493, 1268}

\bibitem[\protect\citeauthoryear{{Birrer} \& {Amara}}{{Birrer} \&
  {Amara}}{2018}]{Birrer2018}
{Birrer} S.,  {Amara} A.,  2018, \mn@doi [Physics of the Dark Universe]
  {10.1016/j.dark.2018.11.002}, \href
  {https://ui.adsabs.harvard.edu/abs/2018PDU....22..189B} {22, 189}

\bibitem[\protect\citeauthoryear{{Birrer}, {Amara}  \& {Refregier}}{{Birrer}
  et~al.}{2017}]{Birrer2017}
{Birrer} S.,  {Amara} A.,   {Refregier} A.,  2017, \mn@doi [\jcap]
  {10.1088/1475-7516/2017/05/037}, \href
  {https://ui.adsabs.harvard.edu/abs/2017JCAP...05..037B} {2017, 037}

\bibitem[\protect\citeauthoryear{{Birrer} et~al.,}{{Birrer}
  et~al.}{2021}]{Birrer2021}
{Birrer} S.,  et~al., 2021, \mn@doi [The Journal of Open Source Software]
  {10.21105/joss.03283}, \href
  {https://ui.adsabs.harvard.edu/abs/2021JOSS....6.3283B} {6, 3283}

\bibitem[\protect\citeauthoryear{{Blandford} \& {Narayan}}{{Blandford} \&
  {Narayan}}{1986}]{BlanfordNarayan86}
{Blandford} R.,  {Narayan} R.,  1986, \mn@doi [\apj] {10.1086/164709}, \href
  {https://ui.adsabs.harvard.edu/abs/1986ApJ...310..568B} {310, 568}

\bibitem[\protect\citeauthoryear{Bond, Cole, Efstathiou  \& Kaiser}{Bond
  et~al.}{1991}]{Bond:1990iw}
Bond J.~R.,  Cole S.,  Efstathiou G.,   Kaiser N.,  1991, \mn@doi [\apj]
  {10.1086/170520}, 379, 440

\bibitem[\protect\citeauthoryear{{Bose}, {Hellwing}, {Frenk}, {Jenkins},
  {Lovell}, {Helly}  \& {Li}}{{Bose} et~al.}{2016}]{Bose++16}
{Bose} S.,  {Hellwing} W.~A.,  {Frenk} C.~S.,  {Jenkins} A.,  {Lovell} M.~R.,
  {Helly} J.~C.,   {Li} B.,  2016, \mn@doi [\mnras] {10.1093/mnras/stv2294},
  \href {https://ui.adsabs.harvard.edu/abs/2016MNRAS.455..318B} {455, 318}

\bibitem[\protect\citeauthoryear{{Bouwens} et~al.,}{{Bouwens}
  et~al.}{2015}]{Bouwens_2015}
{Bouwens} R.~J.,  et~al., 2015, \mn@doi [\apj] {10.1088/0004-637X/803/1/34},
  \href {https://ui.adsabs.harvard.edu/abs/2015ApJ...803...34B} {803, 34}

\bibitem[\protect\citeauthoryear{{Bozek}, {Marsh}, {Silk}  \& {Wyse}}{{Bozek}
  et~al.}{2015}]{Bozek_2015}
{Bozek} B.,  {Marsh} D. J.~E.,  {Silk} J.,   {Wyse} R. F.~G.,  2015, \mn@doi
  [\mnras] {10.1093/mnras/stv624}, \href
  {https://ui.adsabs.harvard.edu/abs/2015MNRAS.450..209B} {450, 209}

\bibitem[\protect\citeauthoryear{{Bullock} \& {Boylan-Kolchin}}{{Bullock} \&
  {Boylan-Kolchin}}{2017}]{Bullock2017}
{Bullock} J.~S.,  {Boylan-Kolchin} M.,  2017, \mn@doi [\araa]
  {10.1146/annurev-astro-091916-055313}, \href
  {https://ui.adsabs.harvard.edu/abs/2017ARA&A..55..343B} {55, 343}

\bibitem[\protect\citeauthoryear{{Bullock}, {Kolatt}, {Sigad}, {Somerville},
  {Kravtsov}, {Klypin}, {Primack}  \& {Dekel}}{{Bullock}
  et~al.}{2001}]{Bullock_2001}
{Bullock} J.~S.,  {Kolatt} T.~S.,  {Sigad} Y.,  {Somerville} R.~S.,  {Kravtsov}
  A.~V.,  {Klypin} A.~A.,  {Primack} J.~R.,   {Dekel} A.,  2001, \mn@doi
  [\mnras] {10.1046/j.1365-8711.2001.04068.x}, \href
  {https://ui.adsabs.harvard.edu/abs/2001MNRAS.321..559B} {321, 559}

\bibitem[\protect\citeauthoryear{{Burkert}}{{Burkert}}{2020}]{Burkert20}
{Burkert} A.,  2020, \mn@doi [\apj] {10.3847/1538-4357/abb242}, \href
  {https://ui.adsabs.harvard.edu/abs/2020ApJ...904..161B} {904, 161}

\bibitem[\protect\citeauthoryear{{Chan}, {Schive}, {Wong}, {Chiueh}  \&
  {Broadhurst}}{{Chan} et~al.}{2020}]{Chan2020}
{Chan} J. H.~H.,  {Schive} H.-Y.,  {Wong} S.-K.,  {Chiueh} T.,   {Broadhurst}
  T.,  2020, \mn@doi [\prl] {10.1103/PhysRevLett.125.111102}, \href
  {https://ui.adsabs.harvard.edu/abs/2020PhRvL.125k1102C} {125, 111102}

\bibitem[\protect\citeauthoryear{{Chan}, {Ferreira}, {May}, {Hayashi}  \&
  {Chiba}}{{Chan} et~al.}{2022}]{Chan22}
{Chan} H. Y.~J.,  {Ferreira} E. G.~M.,  {May} S.,  {Hayashi} K.,   {Chiba} M.,
  2022, \mn@doi [\mnras] {10.1093/mnras/stac063}, \href
  {https://ui.adsabs.harvard.edu/abs/2022MNRAS.511..943C} {511, 943}

\bibitem[\protect\citeauthoryear{{Chen}, {Schive}  \& {Chiueh}}{{Chen}
  et~al.}{2017}]{Chen++2017}
{Chen} S.-R.,  {Schive} H.-Y.,   {Chiueh} T.,  2017, \mn@doi [\mnras]
  {10.1093/mnras/stx449}, \href
  {https://ui.adsabs.harvard.edu/abs/2017MNRAS.468.1338C} {468, 1338}

\bibitem[\protect\citeauthoryear{{Chiba}, {Minezaki}, {Kashikawa}, {Kataza}  \&
  {Inoue}}{{Chiba} et~al.}{2005}]{chiba2005}
{Chiba} M.,  {Minezaki} T.,  {Kashikawa} N.,  {Kataza} H.,   {Inoue} K.~T.,
  2005, \mn@doi [\apj] {10.1086/430403}, \href
  {https://ui.adsabs.harvard.edu/abs/2005ApJ...627...53C} {627, 53}

\bibitem[\protect\citeauthoryear{{Church}, {Mocz}  \& {Ostriker}}{{Church}
  et~al.}{2019}]{Church++19}
{Church} B.~V.,  {Mocz} P.,   {Ostriker} J.~P.,  2019, \mn@doi [\mnras]
  {10.1093/mnras/stz534}, \href
  {https://ui.adsabs.harvard.edu/abs/2019MNRAS.485.2861C} {485, 2861}

\bibitem[\protect\citeauthoryear{{Cicoli}, {Goodsell}  \& {Ringwald}}{{Cicoli}
  et~al.}{2012}]{cicoli2012}
{Cicoli} M.,  {Goodsell} M.~D.,   {Ringwald} A.,  2012, \mn@doi [Journal of
  High Energy Physics] {10.1007/JHEP10(2012)146}, \href
  {https://ui.adsabs.harvard.edu/abs/2012JHEP...10..146C} {2012, 146}

\bibitem[\protect\citeauthoryear{{Cicoli}, {Guidetti}, {Righi}  \&
  {Westphal}}{{Cicoli} et~al.}{2022}]{Cicoli++21}
{Cicoli} M.,  {Guidetti} V.,  {Righi} N.,   {Westphal} A.,  2022, \mn@doi
  [Journal of High Energy Physics] {10.1007/JHEP05(2022)107}, \href
  {https://ui.adsabs.harvard.edu/abs/2022JHEP...05..107C} {2022, 107}

\bibitem[\protect\citeauthoryear{{Cyr-Racine}, {Keeton}  \&
  {Moustakas}}{{Cyr-Racine} et~al.}{2019}]{Cyr-Racine++19}
{Cyr-Racine} F.-Y.,  {Keeton} C.~R.,   {Moustakas} L.~A.,  2019, \mn@doi [\prd]
  {10.1103/PhysRevD.100.023013}, \href
  {https://ui.adsabs.harvard.edu/abs/2019PhRvD.100b3013C} {100, 023013}

\bibitem[\protect\citeauthoryear{{Dalal} \& {Kochanek}}{{Dalal} \&
  {Kochanek}}{2002}]{Dalal2002}
{Dalal} N.,  {Kochanek} C.~S.,  2002, \mn@doi [\apj] {10.1086/340303}, \href
  {https://ui.adsabs.harvard.edu/abs/2002ApJ...572...25D} {572, 25}

\bibitem[\protect\citeauthoryear{{Dalal} \& {Kravtsov}}{{Dalal} \&
  {Kravtsov}}{2022}]{Dalal++22}
{Dalal} N.,  {Kravtsov} A.,  2022, arXiv e-prints, \href
  {https://ui.adsabs.harvard.edu/abs/2022arXiv220305750D} {p. arXiv:2203.05750}

\bibitem[\protect\citeauthoryear{{Dalal}, {Bovy}, {Hui}  \& {Li}}{{Dalal}
  et~al.}{2021}]{Dalal++21}
{Dalal} N.,  {Bovy} J.,  {Hui} L.,   {Li} X.,  2021, \mn@doi [\jcap]
  {10.1088/1475-7516/2021/03/076}, \href
  {https://ui.adsabs.harvard.edu/abs/2021JCAP...03..076D} {2021, 076}

\bibitem[\protect\citeauthoryear{{Davoudiasl} \& {Denton}}{{Davoudiasl} \&
  {Denton}}{2019}]{Davoudiasl_Denton19}
{Davoudiasl} H.,  {Denton} P.~B.,  2019, \mn@doi [\prl]
  {10.1103/PhysRevLett.123.021102}, \href
  {https://ui.adsabs.harvard.edu/abs/2019PhRvL.123b1102D} {123, 021102}

\bibitem[\protect\citeauthoryear{{De Laurentis} \& {Salucci}}{{De Laurentis} \&
  {Salucci}}{2022}]{DeLaurentis_Salucci22}
{De Laurentis} M.,  {Salucci} P.,  2022, \mn@doi [\apj]
  {10.3847/1538-4357/ac54b9}, \href
  {https://ui.adsabs.harvard.edu/abs/2022ApJ...929...17D} {929, 17}

\bibitem[\protect\citeauthoryear{{Della Monica} \& {de Martino}}{{Della Monica}
  \& {de Martino}}{2022}]{DellaMonica++22}
{Della Monica} R.,  {de Martino} I.,  2022, arXiv e-prints, \href
  {https://ui.adsabs.harvard.edu/abs/2022arXiv220603980D} {p. arXiv:2206.03980}

\bibitem[\protect\citeauthoryear{{Despali}, {Giocoli}, {Angulo}, {Tormen},
  {Sheth}, {Baso}  \& {Moscardini}}{{Despali} et~al.}{2016}]{Despali++16}
{Despali} G.,  {Giocoli} C.,  {Angulo} R.~E.,  {Tormen} G.,  {Sheth} R.~K.,
  {Baso} G.,   {Moscardini} L.,  2016, \mn@doi [\mnras]
  {10.1093/mnras/stv2842}, \href
  {https://ui.adsabs.harvard.edu/abs/2016MNRAS.456.2486D} {456, 2486}

\bibitem[\protect\citeauthoryear{{Despali}, {Lovell}, {Vegetti}, {Crain}  \&
  {Oppenheimer}}{{Despali} et~al.}{2020}]{Despali++20}
{Despali} G.,  {Lovell} M.,  {Vegetti} S.,  {Crain} R.~A.,   {Oppenheimer}
  B.~D.,  2020, \mn@doi [\mnras] {10.1093/mnras/stz3068}, \href
  {https://ui.adsabs.harvard.edu/abs/2020MNRAS.491.1295D} {491, 1295}

\bibitem[\protect\citeauthoryear{{Despali}, {Vegetti}, {White}, {Powell},
  {Stacey}, {Fassnacht}, {Rizzo}  \& {Enzi}}{{Despali}
  et~al.}{2022}]{Despali++22}
{Despali} G.,  {Vegetti} S.,  {White} S. D.~M.,  {Powell} D.~M.,  {Stacey}
  H.~R.,  {Fassnacht} C.~D.,  {Rizzo} F.,   {Enzi} W.,  2022, \mn@doi [\mnras]
  {10.1093/mnras/stab3537}, \href
  {https://ui.adsabs.harvard.edu/abs/2022MNRAS.510.2480D} {510, 2480}

\bibitem[\protect\citeauthoryear{{Dhanasingham}, {Cyr-Racine}, {Peter},
  {Benson}  \& {Gilman}}{{Dhanasingham} et~al.}{2022}]{Dhanasingham++22}
{Dhanasingham} B.,  {Cyr-Racine} F.-Y.,  {Peter} A. H.~G.,  {Benson} A.,
  {Gilman} D.,  2022, arXiv e-prints, \href
  {https://ui.adsabs.harvard.edu/abs/2022arXiv220313775D} {p. arXiv:2203.13775}

\bibitem[\protect\citeauthoryear{{Diaz Rivero}, {Cyr-Racine}  \&
  {Dvorkin}}{{Diaz Rivero} et~al.}{2018}]{DiazRivero++18}
{Diaz Rivero} A.,  {Cyr-Racine} F.-Y.,   {Dvorkin} C.,  2018, \mn@doi [\prd]
  {10.1103/PhysRevD.97.023001}, \href
  {https://ui.adsabs.harvard.edu/abs/2018PhRvD..97b3001D} {97, 023001}

\bibitem[\protect\citeauthoryear{{Diemer} \& {Joyce}}{{Diemer} \&
  {Joyce}}{2019}]{DiemerJoyce19}
{Diemer} B.,  {Joyce} M.,  2019, \mn@doi [\apj] {10.3847/1538-4357/aafad6},
  \href {https://ui.adsabs.harvard.edu/abs/2019ApJ...871..168D} {871, 168}

\bibitem[\protect\citeauthoryear{{Dine} \& {Fischler}}{{Dine} \&
  {Fischler}}{1983}]{Dine1982}
{Dine} M.,  {Fischler} W.,  1983, \mn@doi [Physics Letters B]
  {10.1016/0370-2693(83)90639-1}, \href
  {https://ui.adsabs.harvard.edu/abs/1983PhLB..120..137D} {120, 137}

\bibitem[\protect\citeauthoryear{{Du}, {Behrens}  \& {Niemeyer}}{{Du}
  et~al.}{2017}]{Du2016}
{Du} X.,  {Behrens} C.,   {Niemeyer} J.~C.,  2017, \mn@doi [\mnras]
  {10.1093/mnras/stw2724}, \href
  {https://ui.adsabs.harvard.edu/abs/2017MNRAS.465..941D} {465, 941}

\bibitem[\protect\citeauthoryear{{Du}, {Schwabe}, {Niemeyer}  \&
  {B{\"u}rger}}{{Du} et~al.}{2018}]{Du2018}
{Du} X.,  {Schwabe} B.,  {Niemeyer} J.~C.,   {B{\"u}rger} D.,  2018, \mn@doi
  [\prd] {10.1103/PhysRevD.97.063507}, \href
  {https://ui.adsabs.harvard.edu/abs/2018PhRvD..97f3507D} {97, 063507}

\bibitem[\protect\citeauthoryear{{Dutta Chowdhury}, {van den Bosch}, {Robles},
  {van Dokkum}, {Schive}, {Chiueh}  \& {Broadhurst}}{{Dutta Chowdhury}
  et~al.}{2021}]{Chowdhury++21}
{Dutta Chowdhury} D.,  {van den Bosch} F.~C.,  {Robles} V.~H.,  {van Dokkum}
  P.,  {Schive} H.-Y.,  {Chiueh} T.,   {Broadhurst} T.,  2021, \mn@doi [\apj]
  {10.3847/1538-4357/ac043f}, \href
  {https://ui.adsabs.harvard.edu/abs/2021ApJ...916...27D} {916, 27}

\bibitem[\protect\citeauthoryear{{Errani} \& {Pe{\~n}arrubia}}{{Errani} \&
  {Pe{\~n}arrubia}}{2020}]{Errani2019}
{Errani} R.,  {Pe{\~n}arrubia} J.,  2020, \mn@doi [\mnras]
  {10.1093/mnras/stz3349}, \href
  {https://ui.adsabs.harvard.edu/abs/2020MNRAS.491.4591E} {491, 4591}

\bibitem[\protect\citeauthoryear{{Ferreira}}{{Ferreira}}{2021}]{ferreira2021ultralight}
{Ferreira} E. G.~M.,  2021, \mn@doi [\aapr] {10.1007/s00159-021-00135-6}, \href
  {https://ui.adsabs.harvard.edu/abs/2021A&ARv..29....7F} {29, 7}

\bibitem[\protect\citeauthoryear{{Fiacconi}, {Madau}, {Potter}  \&
  {Stadel}}{{Fiacconi} et~al.}{2016}]{Fiacconi++16}
{Fiacconi} D.,  {Madau} P.,  {Potter} D.,   {Stadel} J.,  2016, \mn@doi [\apj]
  {10.3847/0004-637X/824/2/144}, \href
  {https://ui.adsabs.harvard.edu/abs/2016ApJ...824..144F} {824, 144}

\bibitem[\protect\citeauthoryear{{Garrison-Kimmel} et~al.,}{{Garrison-Kimmel}
  et~al.}{2017}]{Garrison-Kimmel++17}
{Garrison-Kimmel} S.,  et~al., 2017, \mn@doi [\mnras] {10.1093/mnras/stx1710},
  \href {https://ui.adsabs.harvard.edu/abs/2017MNRAS.471.1709G} {471, 1709}

\bibitem[\protect\citeauthoryear{{Gavazzi}, {Treu}, {Rhodes}, {Koopmans},
  {Bolton}, {Burles}, {Massey}  \& {Moustakas}}{{Gavazzi}
  et~al.}{2007}]{Gavazzi++07}
{Gavazzi} R.,  {Treu} T.,  {Rhodes} J.~D.,  {Koopmans} L. V.~E.,  {Bolton}
  A.~S.,  {Burles} S.,  {Massey} R.~J.,   {Moustakas} L.~A.,  2007, \mn@doi
  [\apj] {10.1086/519237}, \href
  {https://ui.adsabs.harvard.edu/abs/2007ApJ...667..176G} {667, 176}

\bibitem[\protect\citeauthoryear{{Gilman}, {Agnello}, {Treu}, {Keeton}  \&
  {Nierenberg}}{{Gilman} et~al.}{2017}]{Gilman++17}
{Gilman} D.,  {Agnello} A.,  {Treu} T.,  {Keeton} C.~R.,   {Nierenberg} A.~M.,
  2017, \mn@doi [\mnras] {10.1093/mnras/stx158}, \href
  {https://ui.adsabs.harvard.edu/abs/2017MNRAS.467.3970G} {467, 3970}

\bibitem[\protect\citeauthoryear{{Gilman}, {Birrer}, {Treu}, {Nierenberg}  \&
  {Benson}}{{Gilman} et~al.}{2019}]{Gilman2019c}
{Gilman} D.,  {Birrer} S.,  {Treu} T.,  {Nierenberg} A.,   {Benson} A.,  2019,
  \mn@doi [\mnras] {10.1093/mnras/stz1593}, \href
  {https://ui.adsabs.harvard.edu/abs/2019MNRAS.487.5721G} {487, 5721}

\bibitem[\protect\citeauthoryear{{Gilman}, {Birrer}, {Nierenberg}, {Treu}, {Du}
   \& {Benson}}{{Gilman} et~al.}{2020a}]{Gilman2019a}
{Gilman} D.,  {Birrer} S.,  {Nierenberg} A.,  {Treu} T.,  {Du} X.,   {Benson}
  A.,  2020a, \mn@doi [\mnras] {10.1093/mnras/stz3480}, \href
  {https://ui.adsabs.harvard.edu/abs/2020MNRAS.491.6077G} {491, 6077}

\bibitem[\protect\citeauthoryear{{Gilman}, {Du}, {Benson}, {Birrer},
  {Nierenberg}  \& {Treu}}{{Gilman} et~al.}{2020b}]{Gilman2019b}
{Gilman} D.,  {Du} X.,  {Benson} A.,  {Birrer} S.,  {Nierenberg} A.,   {Treu}
  T.,  2020b, \mn@doi [\mnras] {10.1093/mnrasl/slz173}, \href
  {https://ui.adsabs.harvard.edu/abs/2020MNRAS.492L..12G} {492, L12}

\bibitem[\protect\citeauthoryear{{Gilman}, {Bovy}, {Treu}, {Nierenberg},
  {Birrer}, {Benson}  \& {Sameie}}{{Gilman} et~al.}{2021}]{Gilman2021}
{Gilman} D.,  {Bovy} J.,  {Treu} T.,  {Nierenberg} A.,  {Birrer} S.,  {Benson}
  A.,   {Sameie} O.,  2021, \mn@doi [\mnras] {10.1093/mnras/stab2335}, \href
  {https://ui.adsabs.harvard.edu/abs/2021MNRAS.507.2432G} {507, 2432}

\bibitem[\protect\citeauthoryear{{Gilman}, {Benson}, {Bovy}, {Birrer}, {Treu}
  \& {Nierenberg}}{{Gilman} et~al.}{2022}]{Gilman2022}
{Gilman} D.,  {Benson} A.,  {Bovy} J.,  {Birrer} S.,  {Treu} T.,   {Nierenberg}
  A.,  2022, \mn@doi [\mnras] {10.1093/mnras/stac670}, \href
  {https://ui.adsabs.harvard.edu/abs/2022MNRAS.512.3163G} {512, 3163}

\bibitem[\protect\citeauthoryear{{Glennon}, {Nadler}, {Musoke}, {Banerjee},
  {Prescod-Weinstein}  \& {Wechsler}}{{Glennon} et~al.}{2022}]{Glennon++22}
{Glennon} N.,  {Nadler} E.~O.,  {Musoke} N.,  {Banerjee} A.,
  {Prescod-Weinstein} C.,   {Wechsler} R.~H.,  2022, arXiv e-prints, \href
  {https://ui.adsabs.harvard.edu/abs/2022arXiv220510336G} {p. arXiv:2205.10336}

\bibitem[\protect\citeauthoryear{{Gonz{\'a}lez-Morales}, {Marsh},
  {Pe{\~n}arrubia}  \& {Ure{\~n}a-L{\'o}pez}}{{Gonz{\'a}lez-Morales}
  et~al.}{2017}]{Gonzalez-Morales++17}
{Gonz{\'a}lez-Morales} A.~X.,  {Marsh} D. J.~E.,  {Pe{\~n}arrubia} J.,
  {Ure{\~n}a-L{\'o}pez} L.~A.,  2017, \mn@doi [\mnras] {10.1093/mnras/stx1941},
  \href {https://ui.adsabs.harvard.edu/abs/2017MNRAS.472.1346G} {472, 1346}

\bibitem[\protect\citeauthoryear{{Green}}{{Green}}{2006}]{green2005}
{Green} A.~M.,  2006, in {Manoz} C.,  {Yepes} G.,  eds,  American Institute of
  Physics Conference Series Vol. 878, The Dark Side of the Universe. pp 10--16,
  \mn@doi{10.1063/1.2409062}

\bibitem[\protect\citeauthoryear{{Green} \& {van den Bosch}}{{Green} \& {van
  den Bosch}}{2019}]{green2019}
{Green} S.~B.,  {van den Bosch} F.~C.,  2019, \mn@doi [\mnras]
  {10.1093/mnras/stz2767}, \href
  {https://ui.adsabs.harvard.edu/abs/2019MNRAS.490.2091G} {490, 2091}

\bibitem[\protect\citeauthoryear{{He} et~al.,}{{He} et~al.}{2022}]{He++20}
{He} Q.,  et~al., 2022, \mn@doi [\mnras] {10.1093/mnras/stac191}, \href
  {https://ui.adsabs.harvard.edu/abs/2022MNRAS.511.3046H} {511, 3046}

\bibitem[\protect\citeauthoryear{{Hezaveh}, {Dalal}, {Holder}, {Kisner},
  {Kuhlen}  \& {Perreault Levasseur}}{{Hezaveh} et~al.}{2016a}]{Hezaveh2016}
{Hezaveh} Y.,  {Dalal} N.,  {Holder} G.,  {Kisner} T.,  {Kuhlen} M.,
  {Perreault Levasseur} L.,  2016a, \mn@doi [\jcap]
  {10.1088/1475-7516/2016/11/048}, \href
  {https://ui.adsabs.harvard.edu/abs/2016JCAP...11..048H} {2016, 048}

\bibitem[\protect\citeauthoryear{{Hezaveh}, {Dalal}, {Holder}, {Kisner},
  {Kuhlen}  \& {Perreault Levasseur}}{{Hezaveh} et~al.}{2016b}]{Hezaveh++16}
{Hezaveh} Y.,  {Dalal} N.,  {Holder} G.,  {Kisner} T.,  {Kuhlen} M.,
  {Perreault Levasseur} L.,  2016b, \mn@doi [\jcap]
  {10.1088/1475-7516/2016/11/048}, \href
  {https://ui.adsabs.harvard.edu/abs/2016JCAP...11..048H} {2016, 048}

\bibitem[\protect\citeauthoryear{{Hlo{\v{z}}ek}, {Marsh}  \&
  {Grin}}{{Hlo{\v{z}}ek} et~al.}{2018}]{Hlozek2017}
{Hlo{\v{z}}ek} R.,  {Marsh} D. J.~E.,   {Grin} D.,  2018, \mn@doi [\mnras]
  {10.1093/mnras/sty271}, \href
  {https://ui.adsabs.harvard.edu/abs/2018MNRAS.476.3063H} {476, 3063}

\bibitem[\protect\citeauthoryear{{Hlozek}, {Grin}, {Marsh}  \&
  {Ferreira}}{{Hlozek} et~al.}{2015}]{Hlozek2015}
{Hlozek} R.,  {Grin} D.,  {Marsh} D. J.~E.,   {Ferreira} P.~G.,  2015, \mn@doi
  [\prd] {10.1103/PhysRevD.91.103512}, \href
  {https://ui.adsabs.harvard.edu/abs/2015PhRvD..91j3512H} {91, 103512}

\bibitem[\protect\citeauthoryear{{Hsueh}, {Fassnacht}, {Vegetti}, {McKean},
  {Spingola}, {Auger}, {Koopmans}  \& {Lagattuta}}{{Hsueh}
  et~al.}{2016}]{Hsueh16}
{Hsueh} J.~W.,  {Fassnacht} C.~D.,  {Vegetti} S.,  {McKean} J.~P.,  {Spingola}
  C.,  {Auger} M.~W.,  {Koopmans} L.~V.~E.,   {Lagattuta} D.~J.,  2016, \mn@doi
  [\mnras] {10.1093/mnrasl/slw146}, \href
  {https://ui.adsabs.harvard.edu/abs/2016MNRAS.463L..51H} {463, L51}

\bibitem[\protect\citeauthoryear{{Hsueh} et~al.,}{{Hsueh}
  et~al.}{2017}]{Hsueh17}
{Hsueh} J.~W.,  et~al., 2017, \mn@doi [\mnras] {10.1093/mnras/stx1082}, \href
  {https://ui.adsabs.harvard.edu/abs/2017MNRAS.469.3713H} {469, 3713}

\bibitem[\protect\citeauthoryear{{Hsueh}, {Despali}, {Vegetti}, {Xu},
  {Fassnacht}  \& {Metcalf}}{{Hsueh} et~al.}{2018}]{Hsueh18}
{Hsueh} J.-W.,  {Despali} G.,  {Vegetti} S.,  {Xu} D.,  {Fassnacht} C.~D.,
  {Metcalf} R.~B.,  2018, \mn@doi [\mnras] {10.1093/mnras/stx3320}, \href
  {https://ui.adsabs.harvard.edu/abs/2018MNRAS.475.2438H} {475, 2438}

\bibitem[\protect\citeauthoryear{{Hsueh}, {Enzi}, {Vegetti}, {Auger},
  {Fassnacht}, {Despali}, {Koopmans}  \& {McKean}}{{Hsueh}
  et~al.}{2020}]{Hsueh2019}
{Hsueh} J.~W.,  {Enzi} W.,  {Vegetti} S.,  {Auger} M.~W.,  {Fassnacht} C.~D.,
  {Despali} G.,  {Koopmans} L.~V.~E.,   {McKean} J.~P.,  2020, \mn@doi [\mnras]
  {10.1093/mnras/stz3177}, \href
  {https://ui.adsabs.harvard.edu/abs/2020MNRAS.492.3047H} {492, 3047}

\bibitem[\protect\citeauthoryear{{Hu}, {Barkana}  \& {Gruzinov}}{{Hu}
  et~al.}{2000}]{Hu2000}
{Hu} W.,  {Barkana} R.,   {Gruzinov} A.,  2000, \mn@doi [\prl]
  {10.1103/PhysRevLett.85.1158}, \href
  {https://ui.adsabs.harvard.edu/abs/2000PhRvL..85.1158H} {85, 1158}

\bibitem[\protect\citeauthoryear{{Hui}}{{Hui}}{2021}]{Hui++21}
{Hui} L.,  2021, \mn@doi [\araa] {10.1146/annurev-astro-120920-010024}, \href
  {https://ui.adsabs.harvard.edu/abs/2021ARA&A..59..247H} {59, 247}

\bibitem[\protect\citeauthoryear{Hui, Ostriker, Tremaine  \& Witten}{Hui
  et~al.}{2017}]{Hui2017}
Hui L.,  Ostriker J.~P.,  Tremaine S.,   Witten E.,  2017, \mn@doi [Phys. Rev.
  D] {10.1103/PhysRevD.95.043541}, 95, 043541

\bibitem[\protect\citeauthoryear{{Hui}, {Joyce}, {Landry}  \& {Li}}{{Hui}
  et~al.}{2021}]{Hui2021}
{Hui} L.,  {Joyce} A.,  {Landry} M.~J.,   {Li} X.,  2021, \mn@doi [\jcap]
  {10.1088/1475-7516/2021/01/011}, \href
  {https://ui.adsabs.harvard.edu/abs/2021JCAP...01..011H} {2021, 011}

\bibitem[\protect\citeauthoryear{{Inoue}}{{Inoue}}{2016}]{Inoue16}
{Inoue} K.~T.,  2016, \mn@doi [\mnras] {10.1093/mnras/stw1270}, \href
  {https://ui.adsabs.harvard.edu/abs/2016MNRAS.461..164I} {461, 164}

\bibitem[\protect\citeauthoryear{{Kawai}, {Oguri}, {Amruth}, {Broadhurst}  \&
  {Lim}}{{Kawai} et~al.}{2022}]{kawai2022}
{Kawai} H.,  {Oguri} M.,  {Amruth} A.,  {Broadhurst} T.,   {Lim} J.,  2022,
  \mn@doi [\apj] {10.3847/1538-4357/ac39a2}, \href
  {https://ui.adsabs.harvard.edu/abs/2022ApJ...925...61K} {925, 61}

\bibitem[\protect\citeauthoryear{{Kendall} \& {Easther}}{{Kendall} \&
  {Easther}}{2020}]{Kendall++20}
{Kendall} E.,  {Easther} R.,  2020, \mn@doi [\pasa] {10.1017/pasa.2020.3},
  \href {https://ui.adsabs.harvard.edu/abs/2020PASA...37....9K} {37, e009}

\bibitem[\protect\citeauthoryear{{Kulkarni} \& {Ostriker}}{{Kulkarni} \&
  {Ostriker}}{2022}]{Kulkarni_2021}
{Kulkarni} M.,  {Ostriker} J.~P.,  2022, \mn@doi [\mnras]
  {10.1093/mnras/stab3520}, \href
  {https://ui.adsabs.harvard.edu/abs/2022MNRAS.510.1425K} {510, 1425}

\bibitem[\protect\citeauthoryear{{Lagattuta} et~al.,}{{Lagattuta}
  et~al.}{2010}]{Lagattuta++10}
{Lagattuta} D.~J.,  et~al., 2010, \mn@doi [\apj]
  {10.1088/0004-637X/716/2/1579}, \href
  {https://ui.adsabs.harvard.edu/abs/2010ApJ...716.1579L} {716, 1579}

\bibitem[\protect\citeauthoryear{{Lancaster}, {Giovanetti}, {Mocz}, {Kahn},
  {Lisanti}  \& {Spergel}}{{Lancaster} et~al.}{2020}]{Lancaster++20}
{Lancaster} L.,  {Giovanetti} C.,  {Mocz} P.,  {Kahn} Y.,  {Lisanti} M.,
  {Spergel} D.~N.,  2020, \mn@doi [\jcap] {10.1088/1475-7516/2020/01/001},
  \href {https://ui.adsabs.harvard.edu/abs/2020JCAP...01..001L} {2020, 001}

\bibitem[\protect\citeauthoryear{{Li}, {Hui}  \& {Bryan}}{{Li}
  et~al.}{2019}]{Li++2019}
{Li} X.,  {Hui} L.,   {Bryan} G.~L.,  2019, \mn@doi [\prd]
  {10.1103/PhysRevD.99.063509}, \href
  {https://ui.adsabs.harvard.edu/abs/2019PhRvD..99f3509L} {99, 063509}

\bibitem[\protect\citeauthoryear{{Li}, {Hui}  \& {Yavetz}}{{Li}
  et~al.}{2021}]{Li++2021}
{Li} X.,  {Hui} L.,   {Yavetz} T.~D.,  2021, \mn@doi [\prd]
  {10.1103/PhysRevD.103.023508}, \href
  {https://ui.adsabs.harvard.edu/abs/2021PhRvD.103b3508L} {103, 023508}

\bibitem[\protect\citeauthoryear{{Loken} et~al.,}{{Loken}
  et~al.}{2010}]{Loken2010}
{Loken} C.,  et~al., 2010, in Journal of Physics Conference Series. p. 012026,
  \mn@doi{10.1088/1742-6596/256/1/012026}

\bibitem[\protect\citeauthoryear{Magaña \& Matos}{Magaña \&
  Matos}{2012}]{Magana2012}
Magaña J.,  Matos T.,  2012, \mn@doi [Journal of Physics: Conference Series]
  {10.1088/1742-6596/378/1/012012}, 378, 012012

\bibitem[\protect\citeauthoryear{{Marsh} \& {Pop}}{{Marsh} \&
  {Pop}}{2015}]{MarshPop15}
{Marsh} D. J.~E.,  {Pop} A.-R.,  2015, \mn@doi [\mnras]
  {10.1093/mnras/stv1050}, \href
  {https://ui.adsabs.harvard.edu/abs/2015MNRAS.451.2479M} {451, 2479}

\bibitem[\protect\citeauthoryear{{Mina}, {Mota}  \& {Winther}}{{Mina}
  et~al.}{2020}]{Mina20}
{Mina} M.,  {Mota} D.~F.,   {Winther} H.~A.,  2020, arXiv e-prints, \href
  {https://ui.adsabs.harvard.edu/abs/2020arXiv200704119M} {p. arXiv:2007.04119}

\bibitem[\protect\citeauthoryear{{Minor}, {Kaplinghat}, {Chan}  \&
  {Simon}}{{Minor} et~al.}{2021}]{Minor++21}
{Minor} Q.,  {Kaplinghat} M.,  {Chan} T.~H.,   {Simon} E.,  2021, \mn@doi
  [\mnras] {10.1093/mnras/stab2209}, \href
  {https://ui.adsabs.harvard.edu/abs/2021MNRAS.507.1202M} {507, 1202}

\bibitem[\protect\citeauthoryear{{Mocz} et~al.,}{{Mocz} et~al.}{2019}]{Mocz19}
{Mocz} P.,  et~al., 2019, \mn@doi [\prl] {10.1103/PhysRevLett.123.141301},
  \href {https://ui.adsabs.harvard.edu/abs/2019PhRvL.123n1301M} {123, 141301}

\bibitem[\protect\citeauthoryear{{M{\"u}ller-S{\'a}nchez}, {Prieto}, {Hicks},
  {Vives-Arias}, {Davies}, {Malkan}, {Tacconi}  \&
  {Genzel}}{{M{\"u}ller-S{\'a}nchez} et~al.}{2011}]{MullerSanchez++11}
{M{\"u}ller-S{\'a}nchez} F.,  {Prieto} M.~A.,  {Hicks} E.~K.~S.,  {Vives-Arias}
  H.,  {Davies} R.~I.,  {Malkan} M.,  {Tacconi} L.~J.,   {Genzel} R.,  2011,
  \mn@doi [\apj] {10.1088/0004-637X/739/2/69}, \href
  {https://ui.adsabs.harvard.edu/abs/2011ApJ...739...69M} {739, 69}

\bibitem[\protect\citeauthoryear{{Nadler}, {Birrer}, {Gilman}, {Wechsler},
  {Du}, {Benson}, {Nierenberg}  \& {Treu}}{{Nadler} et~al.}{2021}]{Nadler++21}
{Nadler} E.~O.,  {Birrer} S.,  {Gilman} D.,  {Wechsler} R.~H.,  {Du} X.,
  {Benson} A.,  {Nierenberg} A.~M.,   {Treu} T.,  2021, \mn@doi [\apj]
  {10.3847/1538-4357/abf9a3}, \href
  {https://ui.adsabs.harvard.edu/abs/2021ApJ...917....7N} {917, 7}

\bibitem[\protect\citeauthoryear{{Navarro}, {Frenk}  \& {White}}{{Navarro}
  et~al.}{1997}]{Navarro++97}
{Navarro} J.~F.,  {Frenk} C.~S.,   {White} S. D.~M.,  1997, \mn@doi [\apj]
  {10.1086/304888}, \href
  {https://ui.adsabs.harvard.edu/abs/1997ApJ...490..493N} {490, 493}

\bibitem[\protect\citeauthoryear{{Nierenberg}, {Treu}, {Wright}, {Fassnacht}
  \& {Auger}}{{Nierenberg} et~al.}{2014}]{Nierenberg2014}
{Nierenberg} A.~M.,  {Treu} T.,  {Wright} S.~A.,  {Fassnacht} C.~D.,   {Auger}
  M.~W.,  2014, \mn@doi [\mnras] {10.1093/mnras/stu862}, \href
  {https://ui.adsabs.harvard.edu/abs/2014MNRAS.442.2434N} {442, 2434}

\bibitem[\protect\citeauthoryear{{Nierenberg} et~al.,}{{Nierenberg}
  et~al.}{2017}]{Nierenberg2017}
{Nierenberg} A.~M.,  et~al., 2017, \mn@doi [\mnras] {10.1093/mnras/stx1400},
  \href {https://ui.adsabs.harvard.edu/abs/2017MNRAS.471.2224N} {471, 2224}

\bibitem[\protect\citeauthoryear{{Nierenberg} et~al.,}{{Nierenberg}
  et~al.}{2020}]{Nierenberg2019}
{Nierenberg} A.~M.,  et~al., 2020, \mn@doi [\mnras] {10.1093/mnras/stz3588},
  \href {https://ui.adsabs.harvard.edu/abs/2020MNRAS.492.5314N} {492, 5314}

\bibitem[\protect\citeauthoryear{{Nierenberg} et~al.,}{{Nierenberg}
  et~al.}{2021}]{Nierenberg++21}
{Nierenberg} A.,  et~al., 2021, {A definitive test of the dark matter paradigm
  on small scales}, JWST Proposal. Cycle 1, ID. \#2046

\bibitem[\protect\citeauthoryear{{Nori} \& {Baldi}}{{Nori} \&
  {Baldi}}{2021}]{Nori20}
{Nori} M.,  {Baldi} M.,  2021, \mn@doi [\mnras] {10.1093/mnras/staa3772}, \href
  {https://ui.adsabs.harvard.edu/abs/2021MNRAS.501.1539N} {501, 1539}

\bibitem[\protect\citeauthoryear{{Oguri} \& {Marshall}}{{Oguri} \&
  {Marshall}}{2010}]{oguri2010}
{Oguri} M.,  {Marshall} P.~J.,  2010, \mn@doi [\mnras]
  {10.1111/j.1365-2966.2010.16639.x}, \href
  {https://ui.adsabs.harvard.edu/abs/2010MNRAS.405.2579O} {405, 2579}

\bibitem[\protect\citeauthoryear{{Peccei} \& {Quinn}}{{Peccei} \&
  {Quinn}}{1977}]{peccei1977}
{Peccei} R.~D.,  {Quinn} H.~R.,  1977, \mn@doi [\prl]
  {10.1103/PhysRevLett.38.1440}, \href
  {https://ui.adsabs.harvard.edu/abs/1977PhRvL..38.1440P} {38, 1440}

\bibitem[\protect\citeauthoryear{{Planck Collaboration} et~al.,}{{Planck
  Collaboration} et~al.}{2016}]{Planck2015}
{Planck Collaboration} et~al., 2016, \mn@doi [\aap]
  {10.1051/0004-6361/201527101}, \href
  {https://ui.adsabs.harvard.edu/abs/2016A&A...594A...1P} {594, A1}

\bibitem[\protect\citeauthoryear{{Planck Collaboration} et~al.,}{{Planck
  Collaboration} et~al.}{2020a}]{Planck2018}
{Planck Collaboration} et~al., 2020a, \mn@doi [\aap]
  {10.1051/0004-6361/201833880}, \href
  {https://ui.adsabs.harvard.edu/abs/2020A&A...641A...1P} {641, A1}

\bibitem[\protect\citeauthoryear{{Planck Collaboration} et~al.,}{{Planck
  Collaboration} et~al.}{2020b}]{Planck2020}
{Planck Collaboration} et~al., 2020b, \mn@doi [\aap]
  {10.1051/0004-6361/201833910}, \href
  {https://ui.adsabs.harvard.edu/abs/2020A&A...641A...6P} {641, A6}

\bibitem[\protect\citeauthoryear{{Ponce} et~al.,}{{Ponce}
  et~al.}{2019}]{Ponce2019}
{Ponce} M.,  et~al., 2019, arXiv e-prints, \href
  {https://ui.adsabs.harvard.edu/abs/2019arXiv190713600P} {p. arXiv:1907.13600}

\bibitem[\protect\citeauthoryear{{Preskill}, {Wise}  \& {Wilczek}}{{Preskill}
  et~al.}{1983}]{Preskill1982}
{Preskill} J.,  {Wise} M.~B.,   {Wilczek} F.,  1983, \mn@doi [Physics Letters
  B] {10.1016/0370-2693(83)90637-8}, \href
  {https://ui.adsabs.harvard.edu/abs/1983PhLB..120..127P} {120, 127}

\bibitem[\protect\citeauthoryear{Press \& Schechter}{Press \&
  Schechter}{1974}]{Press:1973iz}
Press W.~H.,  Schechter P.,  1974, \mn@doi [\apj] {10.1086/152650}, 187, 425

\bibitem[\protect\citeauthoryear{{Ritondale}, {Vegetti}, {Despali}, {Auger},
  {Koopmans}  \& {McKean}}{{Ritondale} et~al.}{2019}]{Ritondale++19}
{Ritondale} E.,  {Vegetti} S.,  {Despali} G.,  {Auger} M.~W.,  {Koopmans}
  L.~V.~E.,   {McKean} J.~P.,  2019, \mn@doi [\mnras] {10.1093/mnras/stz464},
  \href {https://ui.adsabs.harvard.edu/abs/2019MNRAS.485.2179R} {485, 2179}

\bibitem[\protect\citeauthoryear{{Rogers} \& {Peiris}}{{Rogers} \&
  {Peiris}}{2021}]{Keir2021}
{Rogers} K.~K.,  {Peiris} H.~V.,  2021, \mn@doi [\prl]
  {10.1103/PhysRevLett.126.071302}, \href
  {https://ui.adsabs.harvard.edu/abs/2021PhRvL.126g1302R} {126, 071302}

\bibitem[\protect\citeauthoryear{Rubin}{Rubin}{1984}]{Rubin1984}
Rubin D.~B.,  1984, The Annals of Statistics, 12, 1151

\bibitem[\protect\citeauthoryear{{Samir Acharya}, {Bobkov}  \& {Kumar}}{{Samir
  Acharya} et~al.}{2010}]{acharya2010}
{Samir Acharya} B.,  {Bobkov} K.,   {Kumar} P.,  2010, \mn@doi [Journal of High
  Energy Physics] {10.1007/JHEP11(2010)105}, \href
  {https://ui.adsabs.harvard.edu/abs/2010JHEP...11..105S} {2010, 105}

\bibitem[\protect\citeauthoryear{{Schive}, {Chiueh}  \& {Broadhurst}}{{Schive}
  et~al.}{2014a}]{Schive2014a}
{Schive} H.-Y.,  {Chiueh} T.,   {Broadhurst} T.,  2014a, \mn@doi [Nature
  Physics] {10.1038/nphys2996}, \href
  {https://ui.adsabs.harvard.edu/abs/2014NatPh..10..496S} {10, 496}

\bibitem[\protect\citeauthoryear{{Schive}, {Liao}, {Woo}, {Wong}, {Chiueh},
  {Broadhurst}  \& {Hwang}}{{Schive} et~al.}{2014b}]{Schive2014b}
{Schive} H.-Y.,  {Liao} M.-H.,  {Woo} T.-P.,  {Wong} S.-K.,  {Chiueh} T.,
  {Broadhurst} T.,   {Hwang} W. Y.~P.,  2014b, \mn@doi [\prl]
  {10.1103/PhysRevLett.113.261302}, \href
  {https://ui.adsabs.harvard.edu/abs/2014PhRvL.113z1302S} {113, 261302}

\bibitem[\protect\citeauthoryear{{Schive}, {Chiueh}, {Broadhurst}  \&
  {Huang}}{{Schive} et~al.}{2016}]{Schive2016}
{Schive} H.-Y.,  {Chiueh} T.,  {Broadhurst} T.,   {Huang} K.-W.,  2016, \mn@doi
  [\apj] {10.3847/0004-637X/818/1/89}, \href
  {https://ui.adsabs.harvard.edu/abs/2016ApJ...818...89S} {818, 89}

\bibitem[\protect\citeauthoryear{{Schneider}}{{Schneider}}{2015}]{Schneider++15}
{Schneider} A.,  2015, \mn@doi [\mnras] {10.1093/mnras/stv1169}, \href
  {https://ui.adsabs.harvard.edu/abs/2015MNRAS.451.3117S} {451, 3117}

\bibitem[\protect\citeauthoryear{{Schutz}}{{Schutz}}{2020}]{Schutz2020}
{Schutz} K.,  2020, \mn@doi [\prd] {10.1103/PhysRevD.101.123026}, \href
  {https://ui.adsabs.harvard.edu/abs/2020PhRvD.101l3026S} {101, 123026}

\bibitem[\protect\citeauthoryear{{Schwabe} \& {Niemeyer}}{{Schwabe} \&
  {Niemeyer}}{2022}]{Schwabe++21}
{Schwabe} B.,  {Niemeyer} J.~C.,  2022, \mn@doi [\prl]
  {10.1103/PhysRevLett.128.181301}, \href
  {https://ui.adsabs.harvard.edu/abs/2022PhRvL.128r1301S} {128, 181301}

\bibitem[\protect\citeauthoryear{{Schwabe}, {Niemeyer}  \& {Engels}}{{Schwabe}
  et~al.}{2016}]{Schwabe16}
{Schwabe} B.,  {Niemeyer} J.~C.,   {Engels} J.~F.,  2016, \mn@doi [\prd]
  {10.1103/PhysRevD.94.043513}, \href
  {https://ui.adsabs.harvard.edu/abs/2016PhRvD..94d3513S} {94, 043513}

\bibitem[\protect\citeauthoryear{{Shajib}, {Treu}, {Birrer}  \&
  {Sonnenfeld}}{{Shajib} et~al.}{2021}]{Shajib++21}
{Shajib} A.~J.,  {Treu} T.,  {Birrer} S.,   {Sonnenfeld} A.,  2021, \mn@doi
  [\mnras] {10.1093/mnras/stab536}, \href
  {https://ui.adsabs.harvard.edu/abs/2021MNRAS.503.2380S} {503, 2380}

\bibitem[\protect\citeauthoryear{{Sheth}, {Mo}  \& {Tormen}}{{Sheth}
  et~al.}{2001}]{sheth2001}
{Sheth} R.~K.,  {Mo} H.~J.,   {Tormen} G.,  2001, \mn@doi [\mnras]
  {10.1046/j.1365-8711.2001.04006.x}, \href
  {https://ui.adsabs.harvard.edu/abs/2001MNRAS.323....1S} {323, 1}

\bibitem[\protect\citeauthoryear{{Spergel}, {Flauger}  \&
  {Hlo{\v{z}}ek}}{{Spergel} et~al.}{2015}]{Spergel_2015}
{Spergel} D.~N.,  {Flauger} R.,   {Hlo{\v{z}}ek} R.,  2015, \mn@doi [\prd]
  {10.1103/PhysRevD.91.023518}, \href
  {https://ui.adsabs.harvard.edu/abs/2015PhRvD..91b3518S} {91, 023518}

\bibitem[\protect\citeauthoryear{{Springel} et~al.,}{{Springel}
  et~al.}{2008}]{Springel++08}
{Springel} V.,  et~al., 2008, \mn@doi [\mnras]
  {10.1111/j.1365-2966.2008.14066.x}, \href
  {https://ui.adsabs.harvard.edu/abs/2008MNRAS.391.1685S} {391, 1685}

\bibitem[\protect\citeauthoryear{{Stacey} \& {McKean}}{{Stacey} \&
  {McKean}}{2018}]{Stacey18}
{Stacey} H.~R.,  {McKean} J.~P.,  2018, \mn@doi [\mnras]
  {10.1093/mnrasl/sly153}, \href
  {https://ui.adsabs.harvard.edu/abs/2018MNRAS.481L..40S} {481, L40}

\bibitem[\protect\citeauthoryear{Su\'arez, Robles  \& Matos}{Su\'arez
  et~al.}{2014}]{Suarez2013}
Su\'arez A.,  Robles V.~H.,   Matos T.,  2014, \mn@doi [Astrophys. Space Sci.
  Proc.] {10.1007/978-3-319-02063-1_9}, 38, 107

\bibitem[\protect\citeauthoryear{{Sugai}, {Kawai}, {Shimono}, {Hattori},
  {Kosugi}, {Kashikawa}, {Inoue}  \& {Chiba}}{{Sugai} et~al.}{2007}]{Sugai07}
{Sugai} H.,  {Kawai} A.,  {Shimono} A.,  {Hattori} T.,  {Kosugi} G.,
  {Kashikawa} N.,  {Inoue} K.~T.,   {Chiba} M.,  2007, \mn@doi [\apj]
  {10.1086/513731}, \href
  {https://ui.adsabs.harvard.edu/abs/2007ApJ...660.1016S} {660, 1016}

\bibitem[\protect\citeauthoryear{{Svrcek} \& {Witten}}{{Svrcek} \&
  {Witten}}{2006}]{Svrcek2006}
{Svrcek} P.,  {Witten} E.,  2006, \mn@doi [Journal of High Energy Physics]
  {10.1088/1126-6708/2006/06/051}, \href
  {https://ui.adsabs.harvard.edu/abs/2006JHEP...06..051S} {2006, 051}

\bibitem[\protect\citeauthoryear{{Treu} et~al.,}{{Treu}
  et~al.}{2018}]{treu2018}
{Treu} T.,  et~al., 2018, \mn@doi [\mnras] {10.1093/mnras/sty2329}, \href
  {https://ui.adsabs.harvard.edu/abs/2018MNRAS.481.1041T} {481, 1041}

\bibitem[\protect\citeauthoryear{{Vegetti}, {Koopmans}, {Auger}, {Treu}  \&
  {Bolton}}{{Vegetti} et~al.}{2014}]{Vegetti2014}
{Vegetti} S.,  {Koopmans} L.~V.~E.,  {Auger} M.~W.,  {Treu} T.,   {Bolton}
  A.~S.,  2014, \mn@doi [\mnras] {10.1093/mnras/stu943}, \href
  {https://ui.adsabs.harvard.edu/abs/2014MNRAS.442.2017V} {442, 2017}

\bibitem[\protect\citeauthoryear{{Vegetti}, {Despali}, {Lovell}  \&
  {Enzi}}{{Vegetti} et~al.}{2018}]{Vegetti++18}
{Vegetti} S.,  {Despali} G.,  {Lovell} M.~R.,   {Enzi} W.,  2018, \mn@doi
  [\mnras] {10.1093/mnras/sty2393}, \href
  {https://ui.adsabs.harvard.edu/abs/2018MNRAS.481.3661V} {481, 3661}

\bibitem[\protect\citeauthoryear{{Wagner-Carena}, {Aalbers}, {Birrer},
  {Nadler}, {Darragh-Ford}, {Marshall}  \& {Wechsler}}{{Wagner-Carena}
  et~al.}{2022}]{Wagner-Carena++22}
{Wagner-Carena} S.,  {Aalbers} J.,  {Birrer} S.,  {Nadler} E.~O.,
  {Darragh-Ford} E.,  {Marshall} P.~J.,   {Wechsler} R.~H.,  2022, arXiv
  e-prints, \href {https://ui.adsabs.harvard.edu/abs/2022arXiv220300690W} {p.
  arXiv:2203.00690}

\bibitem[\protect\citeauthoryear{{Walker}, {Mateo}, {Olszewski}, {Gnedin},
  {Wang}, {Sen}  \& {Woodroofe}}{{Walker} et~al.}{2007}]{Walker++07}
{Walker} M.~G.,  {Mateo} M.,  {Olszewski} E.~W.,  {Gnedin} O.~Y.,  {Wang} X.,
  {Sen} B.,   {Woodroofe} M.,  2007, \mn@doi [\apjl] {10.1086/521998}, \href
  {https://ui.adsabs.harvard.edu/abs/2007ApJ...667L..53W} {667, L53}

\bibitem[\protect\citeauthoryear{{Walker}, {Mateo}, {Olszewski},
  {Pe{\~n}arrubia}, {Evans}  \& {Gilmore}}{{Walker} et~al.}{2009}]{Walker++09}
{Walker} M.~G.,  {Mateo} M.,  {Olszewski} E.~W.,  {Pe{\~n}arrubia} J.,  {Evans}
  N.~W.,   {Gilmore} G.,  2009, \mn@doi [\apj] {10.1088/0004-637X/704/2/1274},
  \href {https://ui.adsabs.harvard.edu/abs/2009ApJ...704.1274W} {704, 1274}

\bibitem[\protect\citeauthoryear{{Wang}, {Mao}, {Zentner}, {Lange}, {van den
  Bosch}  \& {Wechsler}}{{Wang} et~al.}{2020}]{Wang++20}
{Wang} K.,  {Mao} Y.-Y.,  {Zentner} A.~R.,  {Lange} J.~U.,  {van den Bosch}
  F.~C.,   {Wechsler} R.~H.,  2020, \mn@doi [\mnras] {10.1093/mnras/staa2733},
  \href {https://ui.adsabs.harvard.edu/abs/2020MNRAS.498.4450W} {498, 4450}

\bibitem[\protect\citeauthoryear{{Webb} \& {Bovy}}{{Webb} \&
  {Bovy}}{2020}]{WebbBovy20}
{Webb} J.~J.,  {Bovy} J.,  2020, \mn@doi [\mnras] {10.1093/mnras/staa2852},
  \href {https://ui.adsabs.harvard.edu/abs/2020MNRAS.499..116W} {499, 116}

\bibitem[\protect\citeauthoryear{{Wechsler}, {Bullock}, {Primack}, {Kravtsov}
  \& {Dekel}}{{Wechsler} et~al.}{2002}]{Wechsler_2002}
{Wechsler} R.~H.,  {Bullock} J.~S.,  {Primack} J.~R.,  {Kravtsov} A.~V.,
  {Dekel} A.,  2002, \mn@doi [\apj] {10.1086/338765}, \href
  {https://ui.adsabs.harvard.edu/abs/2002ApJ...568...52W} {568, 52}

\bibitem[\protect\citeauthoryear{{Weinberg}}{{Weinberg}}{1978}]{weinberg1978}
{Weinberg} S.,  1978, \mn@doi [\prl] {10.1103/PhysRevLett.40.223}, \href
  {https://ui.adsabs.harvard.edu/abs/1978PhRvL..40..223W} {40, 223}

\bibitem[\protect\citeauthoryear{{Wilczek}}{{Wilczek}}{1978}]{wilczek1978}
{Wilczek} F.,  1978, \mn@doi [\prl] {10.1103/PhysRevLett.40.279}, \href
  {https://ui.adsabs.harvard.edu/abs/1978PhRvL..40..279W} {40, 279}

\bibitem[\protect\citeauthoryear{{Witten}}{{Witten}}{1984}]{Witten1984}
{Witten} E.,  1984, \mn@doi [Physics Letters B] {10.1016/0370-2693(84)90422-2},
  \href {https://ui.adsabs.harvard.edu/abs/1984PhLB..149..351W} {149, 351}

\bibitem[\protect\citeauthoryear{{Yavetz}, {Li}  \& {Hui}}{{Yavetz}
  et~al.}{2022}]{Yavetz++22}
{Yavetz} T.~D.,  {Li} X.,   {Hui} L.,  2022, \mn@doi [\prd]
  {10.1103/PhysRevD.105.023512}, \href
  {https://ui.adsabs.harvard.edu/abs/2022PhRvD.105b3512Y} {105, 023512}

\bibitem[\protect\citeauthoryear{{Zelko}, {Treu}, {Abazajian}, {Gilman},
  {Benson}, {Birrer}, {Nierenberg}  \& {Kusenko}}{{Zelko}
  et~al.}{2022}]{Zelko++22}
{Zelko} I.~A.,  {Treu} T.,  {Abazajian} K.~N.,  {Gilman} D.,  {Benson} A.~J.,
  {Birrer} S.,  {Nierenberg} A.~M.,   {Kusenko} A.,  2022, arXiv e-prints,
  \href {https://ui.adsabs.harvard.edu/abs/2022arXiv220509777Z} {p.
  arXiv:2205.09777}

\makeatother
\end{thebibliography}

\bsp	% typesetting comment
\label{lastpage}
\end{document}